%% file: main.tex
\documentclass[conference]{IEEEtran}
\IEEEoverridecommandlockouts

\PassOptionsToPackage{table,xcdraw}{xcolor}
\usepackage{colortbl}
\usepackage{tcolorbox}
\usepackage{tikz}
\usepackage{mathdots}
\usetikzlibrary{fadings}
\usetikzlibrary{patterns}
\usetikzlibrary{shadows.blur}
\usetikzlibrary{shapes}
\usetikzlibrary{shapes.geometric, arrows.meta, positioning}
\usepackage{listings}
\usepackage{array}
\usepackage{siunitx}
\usepackage{multirow}
\usepackage{tabularx}
\usepackage{extarrows}
\usepackage{booktabs}
\usepackage{enumitem}
\usepackage{titlesec}
\usepackage{caption}
\usepackage{textcomp}
\usepackage{url}

\usepackage{cite}
\usepackage{amsmath,amssymb,amsfonts}
\usepackage{algorithmic}
\usepackage{graphicx}
\usepackage{textcomp}
\usepackage{xcolor}
\def\BibTeX{{\rm B\kern-.05em{\sc i\kern-.025em b}\kern-.08em
    T\kern-.1667em\lower.7ex\hbox{E}\kern-.125emX}}

\newcommand{\tool}[0]{\mbox{\textsc{SlimStart}}}

\newcommand{\lstbg}[3][0pt]{{\fboxsep#1\colorbox{#2}{\strut #3}}}
\lstdefinelanguage{diff}{
  basicstyle=\footnotesize \ttfamily \color{black},
  columns=fullflexible,
  breaklines=true,
  breakatwhitespace=false,
  showspaces=false,               
  showstringspaces=false,  
  frame=single, 
  showtabs=false,
  numbersep=5pt,
  showstringspaces=false,        
  stepnumber=1,                   
  tabsize=5,                     
  title=\lstname,  
  numbers=left,                 
  numbersep=5pt,   
  backgroundcolor=\color{white},
  morecomment=[f][\lstbg{red!30}]-,
  morecomment=[f][\lstbg{green!30}]+,
  morecomment=[f][\textit]{@@},
  moredelim=**[is][\color{red}]{@}{@},
  moredelim=**[is][\color{green}]{##}{##},
}
\definecolor{white}{rgb}{0.98,0.98,0.98}
\definecolor{dkgreen}{rgb}{0,0.6,0}
\definecolor{dred}{rgb}{0.545,0,0}
\definecolor{dblue}{rgb}{0,0,0.545}
\definecolor{lgrey}{rgb}{255,0.9,0.9}
\definecolor{gray}{rgb}{0.4,0.4,0.4}
\definecolor{darkblue}{rgb}{0.0,0.0,0.6}

\definecolor{pblue}{rgb}{0.13,0.13,1}
\definecolor{pgreen}{rgb}{0,0.5,0}
\definecolor{pred}{rgb}{0.9,0,0}
\definecolor{pgrey}{rgb}{0.46,0.45,0.48}
\lstdefinestyle{prStyle}{
  showspaces=false,
  showtabs=false,
  breaklines=true,
  showstringspaces=false,
  breakatwhitespace=true,
  postbreak=\mbox{\textcolor{black}{$\hookrightarrow$}\space},
  captionpos=b,    
  commentstyle=\color{pgreen},
  keywordstyle=\color{pblue},
  stringstyle=\color{pred},
  basicstyle=\ttfamily,
  moredelim=[il][\textcolor{pgrey}]{},
  frame=tb,
  moredelim=[is][\textcolor{pgrey}]{\%\%}{\%\%}
  moredelim=**[is][\color{red}]{@}{@},
  moredelim=**[is][\color{dkgreen}]{##}{##},
}
\lstdefinelanguage{cpp}{
  backgroundcolor=\color{white},  
  basicstyle=\scriptsize \ttfamily \color{black} ,   
  breakatwhitespace=false,       
  breaklines=true,
  postbreak=\mbox{\textcolor{black}{$\hookrightarrow$}\space},
  captionpos=b,                   
  commentstyle=\color{dkgreen},   
  deletekeywords={...},          
  escapeinside={\%*}{*)},                  
  frame=single,                  
  language=C++,                
  keywordstyle=\color{dblue},  
  morekeywords={BRIEFDescriptorConfig,string,TiXmlNode,DetectorDescriptorConfigContainer,istringstream,cerr,exit}, 
  identifierstyle=\color{black},
  stringstyle=\color{blue},      
  numbers=right,                 
  numbersep=5pt,                  
  numberstyle=\tiny\color{black}, 
  rulecolor=\color{black},        
  showspaces=false,               
  showstringspaces=false,        
  showtabs=false,                
  stepnumber=1,                   
  tabsize=5,                     
  title=\lstname,
  moredelim=**[is][\color{red}]{@}{@},
  moredelim=**[is][\color{dkgreen}]{##}{##},
}

\usepackage{amsthm}
\newtheorem{observation}{Observation}

\newcommand\copyrighttext{%
  \footnotesize \textcopyright 2025 IEEE. Personal use of this material is permitted.
  Permission from IEEE must be obtained for all other uses, in any current or future
  media, including reprinting/republishing this material for advertising or promotional
  purposes, creating new collective works, for resale or redistribution to servers or
  lists, or reuse of any copyrighted component of this work in other works.
  }
\newcommand\copyrightnotice{%
\begin{tikzpicture}[remember picture,overlay]
\node[anchor=south,yshift=10pt] at (current page.south) {\fbox{\parbox{\dimexpr\textwidth-\fboxsep-\fboxrule\relax}{\copyrighttext}}};
\end{tikzpicture}%
}

\begin{document}

\title{
Efficient Serverless Cold Start: Reducing Library Loading Overhead by Profile-guided Optimization
}

\author{\IEEEauthorblockN{Syed Salauddin Mohammad Tariq\textsuperscript{*}, Ali Al Zein\textsuperscript{*}, Soumya Sripad Vaidya, Arati Khanolkar, Zheng Song, Probir Roy}
\IEEEauthorblockA{\textit{University of Michigan Dearborn}, Dearborn, Michigan, USA\\
\{ssmtariq, alielzei, soumyasv, aratik, zhesong, probirr\}@umich.edu}

\thanks{\textsuperscript{*}The authors have equal contribution.

This material is partly based on work supported by the National Science Foundation award CNS-2006373

This paper has been accepted for publication at the 45th IEEE International Conference on Distributed Computing Systems (ICDCS 2025). This is the preprint version.

}
}

\maketitle
\copyrightnotice

\begin{abstract}

Serverless computing abstracts away server management, enabling automatic scaling, efficient resource utilization, and cost-effective pricing models. However, despite these advantages, it faces the significant challenge of cold-start latency, adversely impacting end-to-end performance. Our study shows that many serverless functions initialize libraries that are rarely or never used under typical workloads, thus introducing unnecessary overhead. Although existing static analysis techniques can identify unreachable libraries, they fail to address workload-dependent inefficiencies, resulting in limited performance improvements. To overcome these limitations, we present \tool{}, a profile-guided optimization tool designed to identify and mitigate inefficient library usage patterns in serverless applications. By leveraging statistical sampling and call-path profiling, \tool{} collects runtime library usage data, generates detailed optimization reports, and applies automated code transformations to reduce cold-start overhead.  Furthermore, \tool{} integrates seamlessly into CI/CD pipelines, enabling adaptive monitoring and continuous optimizations tailored to evolving workloads.  Through extensive evaluation across three benchmark suites and four real-world serverless applications, \tool{} achieves up to a 2.30$\times$ speedup in initialization latency, a 2.26$\times$ improvement in end-to-end latency, and a 1.51$\times$ reduction in memory usage, demonstrating its effectiveness in addressing cold-start inefficiencies and optimizing resource utilization.
\end{abstract}


\input{sections/1-introduction}

\input{sections/2-background}

\input{sections/2.1-inefficiency-categorization}

\input{sections/5-Design-implementation}
\input{sections/6-Evaluation}

\input{sections/7-Case-studies}
\input{sections/3-Related-work}
\input{sections/9-Conclusions}
\input{sections/11-references}
\end{document}

%% file: sections/1-introduction.tex
\section{Introduction}
\label{Intro}
Serverless computing has emerged as a popular paradigm that simplifies cloud application development and deployment by abstracting the infrastructure management complexities. However, similar to other cloud computing paradigms, serverless computing must optimize \textbf{end-to-end latency}~\cite{yang2022infless}—the time from receiving a request to completing it. End-to-end latency is closely linked to user satisfaction and company revenue. For example, Amazon reported a 1\% drop in sales for every 100ms increase in latency\cite{amazon_study}, and Google found that a 500ms delay caused a 20\% drop in traffic~\cite{google_study}.

However, achieving low end-to-end latency in serverless computing is often hindered by the challenge of \textbf{cold-start latency}~\cite{shahrad2020serverless, wang2018peeking}. When serverless functions remain idle beyond a predefined \textit{keep-alive} period, their resources are reclaimed for other workloads. As a result, invoking the function again requires a time-consuming re-initialization process, commonly referred to as a cold start, which significantly increases overall end-to-end latency.

To mitigate cold-start latency, many serverless computing advancements have primarily focused on optimizing the runtime systems of serverless hosting platforms. Such approaches include shared resource utilization~\cite{li2022help}, automatic memory deduplication~\cite{saxena2022memory}, function caching~\cite{chen2023s}, function compression~\cite{basu2024codecrunch}, and the reuse of pre-warmed instances~\cite{bhasi2021kraken, gunasekaran2020fifer, roy2022icebreaker, shahrad2020serverless}. 

As a complementary strategy to runtime optimization, recent research~\cite{liu2023faaslight} highlights a promising yet underexplored approach to reducing serverless function overhead: optimizing the applications themselves, particularly the library initialization process. While serverless functions are often small, typically comprising only a few hundred lines of code, they frequently rely on external libraries for core functionality. These libraries, however, are often substantial—containing hundreds or thousands of lines of code—and were not originally designed with serverless environments in mind. In many cases, a significant portion of cold-start time is wasted loading libraries or modules that are unnecessary for the function's execution~\cite{yu2024rainbowcake,liu2023faaslight}.

To reduce the cold-start latency caused by superfluous libraries, the state-of-the-art solution~\cite{liu2023faaslight} employed static analysis to identify and eliminate libraries that are unreachable from any serverless entry function. However, our empirical study reveals that even libraries deemed reachable are often rarely or never invoked in real-world serverless workloads. Our investigation shows that many serverless functions have multiple entry points, but only a small subset of these are frequently triggered under typical workloads. Libraries linked to entry points that are specific to certain workloads, rather than being universally required, represent \textbf{workload-dependent libraries}. Static analysis cannot detect such libraries because it evaluates code reachability without considering runtime workload dynamics. This highlights the need for a runtime profiling approach.

To fill this gap, this paper introduces \tool{}, a profile-guided optimization tool designed to identify workload-dependent superfluous libraries and optimize their loading automatically. \tool{} consists of three main components: an attachable runtime profiler that integrates seamlessly into production cloud environments, an automated code optimizer for lazy loading superfluous libraries, and an adaptive mechanism that dynamically monitors workload
variations and triggers the profiler and the optimizer. During serverless application invocations, the profiler gathers detailed data on library usage and their corresponding calling contexts, enabling targeted and effective optimizations.

Implementing a profile-guided optimization tool for serverless environments involves addressing several technical challenges. First, monitoring library usage in latency-sensitive applications requires a lightweight approach to minimize performance overhead, as traditional methods like binary or source code instrumentation are too resource-intensive. Second, accurately measuring library utilization is complicated by cascading dependencies, where libraries may indirectly invoke others, making it difficult to attribute actual usage. 

To overcome these challenges, \tool{} employs lightweight profiling and innovative techniques to accurately capture library usage patterns in real-world serverless workloads.

While we implement \tool{} for Python-based serverless functions, as 58\% of serverless applications are written in Python, making it the dominant language in serverless development~\cite{datadog2021state}, the solution is adaptable for other programming languages as well. \tool{} is integrated seamlessly with CI/CD pipelines, where it automates the periodic profiling, detection, and optimization of inefficient library usage as part of the deployment workflow. We perform an extensive evaluation to verify the effectiveness of \tool{} using three Python-based serverless application benchmarks and four real-world serverless applications. Our evaluation demonstrates that \tool{} achieves up to 2.30$\times$ speedup in initialization latency, 2.26$\times$ speedup in end-to-end latency, and a 1.51$\times$ reduction in memory usage. \tool{} demonstrates superior performance improvements by dynamically identifying and optimizing workload-dependent library usage, achieving an average of 14.29\% better end-to-end latency reduction and 27.72\% better memory reduction compared to the state-of-the-art static analysis-based methods~\cite{liu2023faaslight}.

\textbf{Paper Contributions}
In summary, this paper makes the following  contributions:

\begin{itemize} [leftmargin=*]
\item This paper is the first to conduct an in-depth study revealing that serverless applications often initialize libraries that are rarely or never used due to uneven workload distribution. The unnecessary initialization of these workload-dependent libraries introduces significant overhead during cold starts, and the study highlights how static analysis misses opportunities to detect and optimize these inefficiencies.

\item We implement \tool{}, a profile-guided code optimization tool that identifies and optimizes inefficient library usage patterns in serverless applications. \tool{} employs innovative techniques to reduce profiling overhead while accurately attributing and optimizing inefficient library usage. \tool{} seamlessly integrates into CI/CD pipelines, using data-driven adaptive monitoring to continuously track evolving serverless workloads and apply optimizations, thus ensuring continuous performance improvements.

\item This paper extensively evaluates \tool{} on three benchmark suites and four real-world serverless applications. The results demonstrate up to 2.30$\times$ speedup in initialization latency, 2.26$\times$ speedup in end-to-end latency, and a 1.51$\times$ reduction in memory usage.
\end{itemize}

The rest of the paper is organized as follows: Section~\ref{background} provides the empirical study, motivating the need for workload-aware code optimization to reduce cold-start latency. Section~\ref{design} presents the overview of \tool{} design and technical challenges. Section~\ref{methodology} details the methodology and implementation of \tool{}. In Section~\ref{eval}, we evaluate \tool{} and present case studies. Section~\ref{relatedwork} discusses the related work of \tool{}. Finally, Section~\ref{conclusions} concludes the paper.

%% file: sections/2-background.tex
\section{Motivation}
\label{background}
\label{motive}
This section introduces our empirical study and its results, which motivated our work. 

\subsection{Optimizing Library Loading: Why and How}
\begin{figure}[!hbt]
    \includegraphics[width=\linewidth]{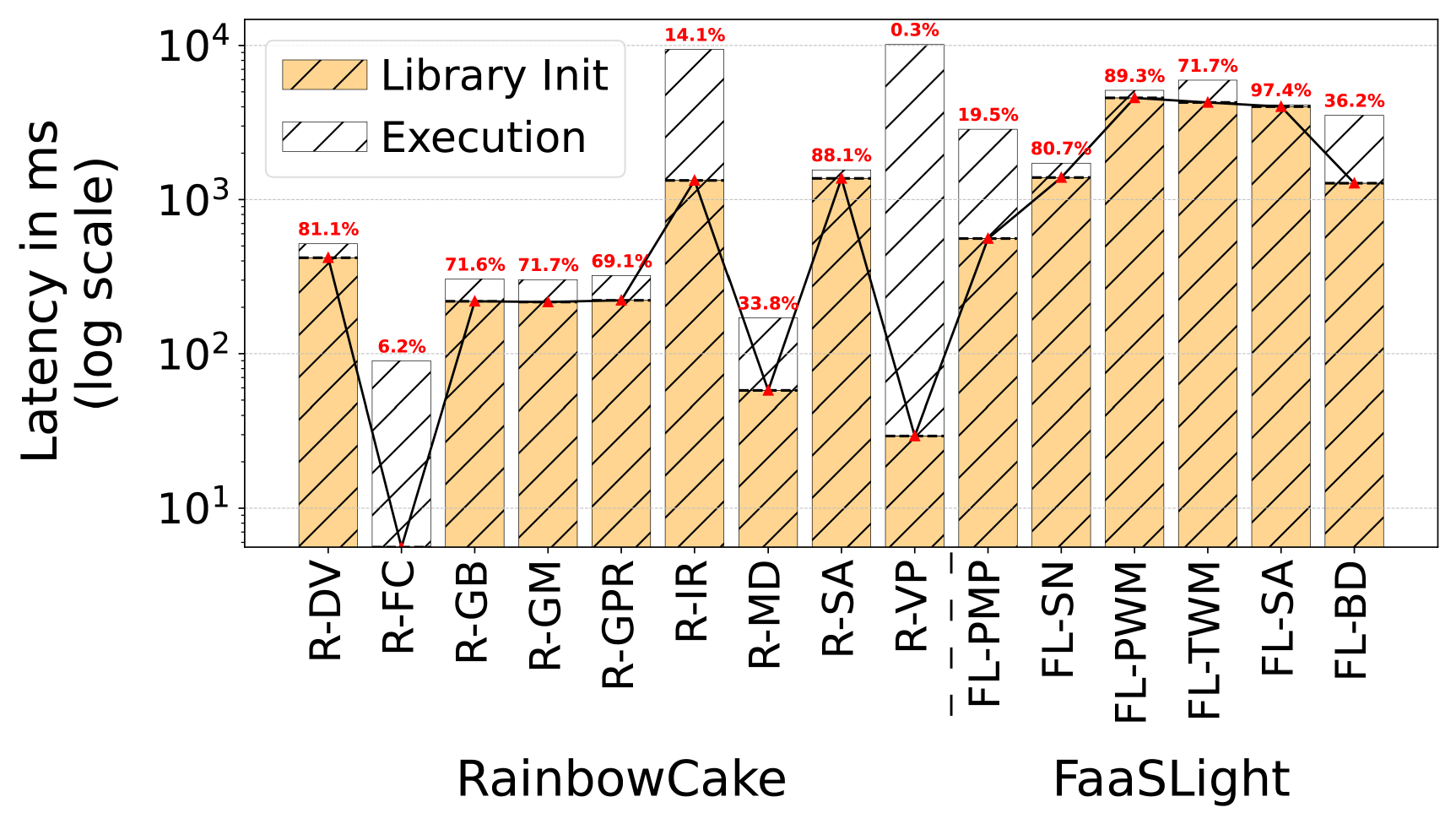}
    \caption{Ratio of library Initialization time to end-to-end time.}
    \label{fig:init_to_e2e}
\end{figure}
To quantify the impact of library initialization on overall end-to-end time, we evaluated a collection of serverless Python applications drawn from existing literature~\cite{yu2024rainbowcake,liu2023faaslight}. Figure \ref{fig:init_to_e2e} presents the library initialization time, end-to-end time, and their respective ratios. The results demonstrate that, for the majority of serverless applications, library initialization contributes to more than 70\% of the total end-to-end time. These findings highlight the critical importance of optimizing library initialization to significantly reduce cold-start latency in serverless Python applications.

\begin{observation}
Library initialization introduces considerable overhead to end-to-end latency during a cold start.
\end{observation}

\begin{table}[!htbp]
    \setlength{\belowcaptionskip}{-3pt}
    \centering
    \small
    \begin{tabular}{>{\raggedright\arraybackslash}p{7.5cm}}
        \hline
        \rowcolor{gray!30}\textbf{File: igraph/clustering.py, Lines 11-13} \\
        \hline
        \texttt{from igraph.drawing.colors import ...} \\
        \texttt{from igraph.drawing.cairo.dendrogram import ...} \\
        \texttt{from igraph.drawing.matplotlib.dendrogram import ...} \\
        \hline
        \rowcolor{gray!30}\textbf{Call Path} \\
        \hline
        handler.py:2 \\
        \quad$\rightarrow$ igraph/\_\_init\_\_.py:104 \\
        \quad\quad$\rightarrow$ igraph/community.py:2 \\
        \quad\quad\quad$\rightarrow$ igraph/clustering.py:11-13 \\
        \hline
    \end{tabular}
    \caption{Importing unused libraries in \textit{graph\_bfs}.}
    \label{tab:motiv_example_unused_lib}
\end{table}
Furthermore, we conducted a manual analysis of 22 serverless applications drawn from three benchmark suites (RainbowCake~\cite{yu2024rainbowcake}, Faaslight~\cite{liu2023faaslight}, and FaaSWorkbench~\cite{kim2019functionbench}) and four real-world serverless applications (CVE-bin-tool~\cite{intel2024cvebin}, OCRmyPDF~\cite{ocrmypdf2024github}, Sensor telemetry data analysis~\cite{constable2024environmental}, and Heart failure prediction~\cite{sakhiya2024heart}) to examine library usage patterns. Our analysis revealed that 17 serverless applications unnecessarily initialize non-essential library modules, leading to significant overheads during cold starts. Table~\ref{tab:motiv_example_unused_lib} showcases a code snippet from the RainbowCake-\textit{graph\_bfs} serverless application, highlighting its dependency on the \textit{igraph} library as a motivating example. The \textit{graph\_bfs} application executes a breadth-first search on a generated graph using the \textit{igraph} library. \textit{igraph} is a comprehensive toolset for graph analysis, including robust graph visualization capabilities. When imported by the \textit{graph\_bfs} application, \textit{igraph} initializes many features by default, including its visualization tools. However, the \textit{graph\_bfs} application only utilizes \textit{igraph} for graph traversal, making the initialization of the visualization capabilities unnecessary. Our experiments indicate that \textit{igraph}'s visualization tool contributes to a 37\% overhead in initialization time for the \textit{graph\_bfs} application. By manually disabling the initialization of the visualization tool and other non-essential components, we achieved a 1.65$\times$ improvement in the library's initialization time for the application.

\subsection{Deficiency of Static-Analysis based Detection}
\label{staticvsDynamic}
Although our analysis reveals the prevalence of unnecessary library usage in serverless applications, identifying these libraries is a challenging task due to the extensive codebases of libraries and their numerous modules. To address this challenge, prior work such as FaaSLight~\cite{liu2023faaslight} explored static analysis-based methods, leveraging reachability analysis to identify unused libraries and modules systematically. FaaSLight reported an average latency reduction of 19.21\% by eliminating these libraries. 

However, FaaSLight does not explicitly demonstrate whether or to what extent it effectively avoids loading superfluous libraries. To investigate this, we deployed five serverless applications referenced in the FaaSLight paper and analyzed their library usage patterns during execution over a period of 1 week. To estimate the upper-bound latency reduction achievable by optimizing library loading, we employed a dynamic profiling approach using statistical sampling. This method periodically captures library usage during execution by collecting a large number of samples. Leveraging the law of large numbers, libraries with no observed samples across sufficient executions are confidently deemed unused. By identifying unused libraries and estimating the time saved by avoiding their initialization, this approach approximates the maximum latency improvements achievable by lazy loading, addressing limitations of static analysis techniques like those used in FaaSLight.


\begin{figure}
    \setlength{\belowcaptionskip}{-10pt}
    \includegraphics[width=\linewidth]{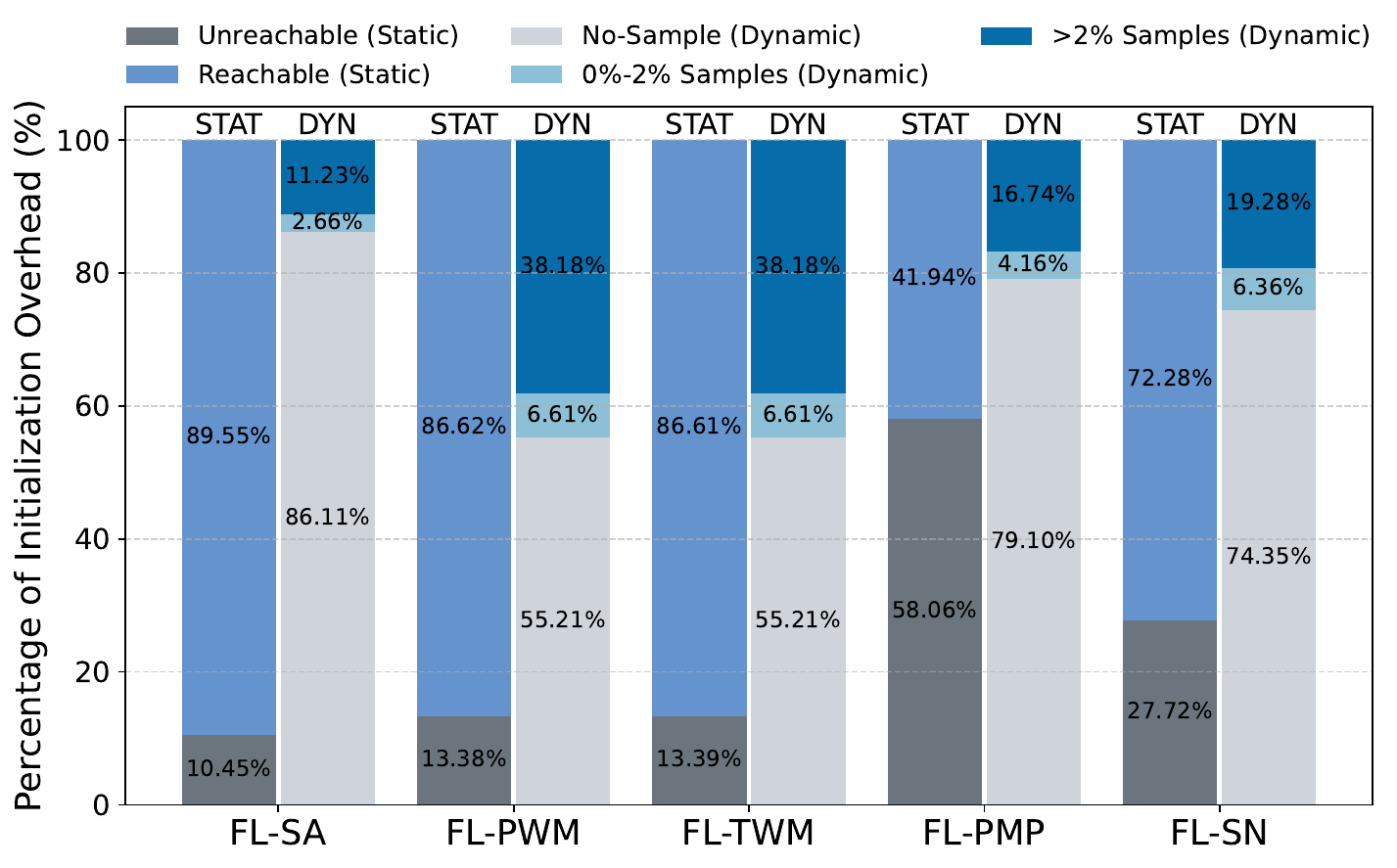}
    \caption{Comparison of library initialization overhead in serverless applications, grouped by library categorization from static reachability analysis (STAT) and dynamic profiling via statistical sampling (DYN). The dynamic profiling highlights finer-grained insights into library usage during execution, emphasizing the differences in overhead contributions.}
    \label{fig:reachabilityVsSampling}
\end{figure}


Figure~\ref{fig:reachabilityVsSampling} compares the latency optimization achieved by FaaSLight with the upper bound estimated through dynamic profiling. This upper bound reflects the initialization overhead of libraries that are either not sampled or rarely sampled (less than 2\% of samples). Dynamic profiling reveals that these libraries contribute significantly to unnecessary initialization overhead, highlighting optimization opportunities overlooked by static analysis. Unlike static methods that treat all reachable libraries as essential, dynamic profiling provides a more accurate assessment by capturing real workload patterns. This analysis shows that accounting for workload-dependent libraries can achieve latency reductions averaging 50.68\%, with a range from 25.2\% for FL-PMP to 78.32\% for FL-SA. These results emphasize the limitations of static analysis and the advantages of workload-aware profiling. Based on the results, we make the following observation: 

\begin{observation}
The static-analysis approach for detecting superfluous libraries suffers from a high false-positive rate — many libraries it identifies as reachable are, in fact, unreachable in a typical workload. This presents an opportunity to further reduce latency by eliminating false positives.
\end{observation}


\subsection{Potential Impact of Addressing such Deficiency}
\begin{figure}
    \setlength{\belowcaptionskip}{-15pt}
    \includegraphics[width=\linewidth]{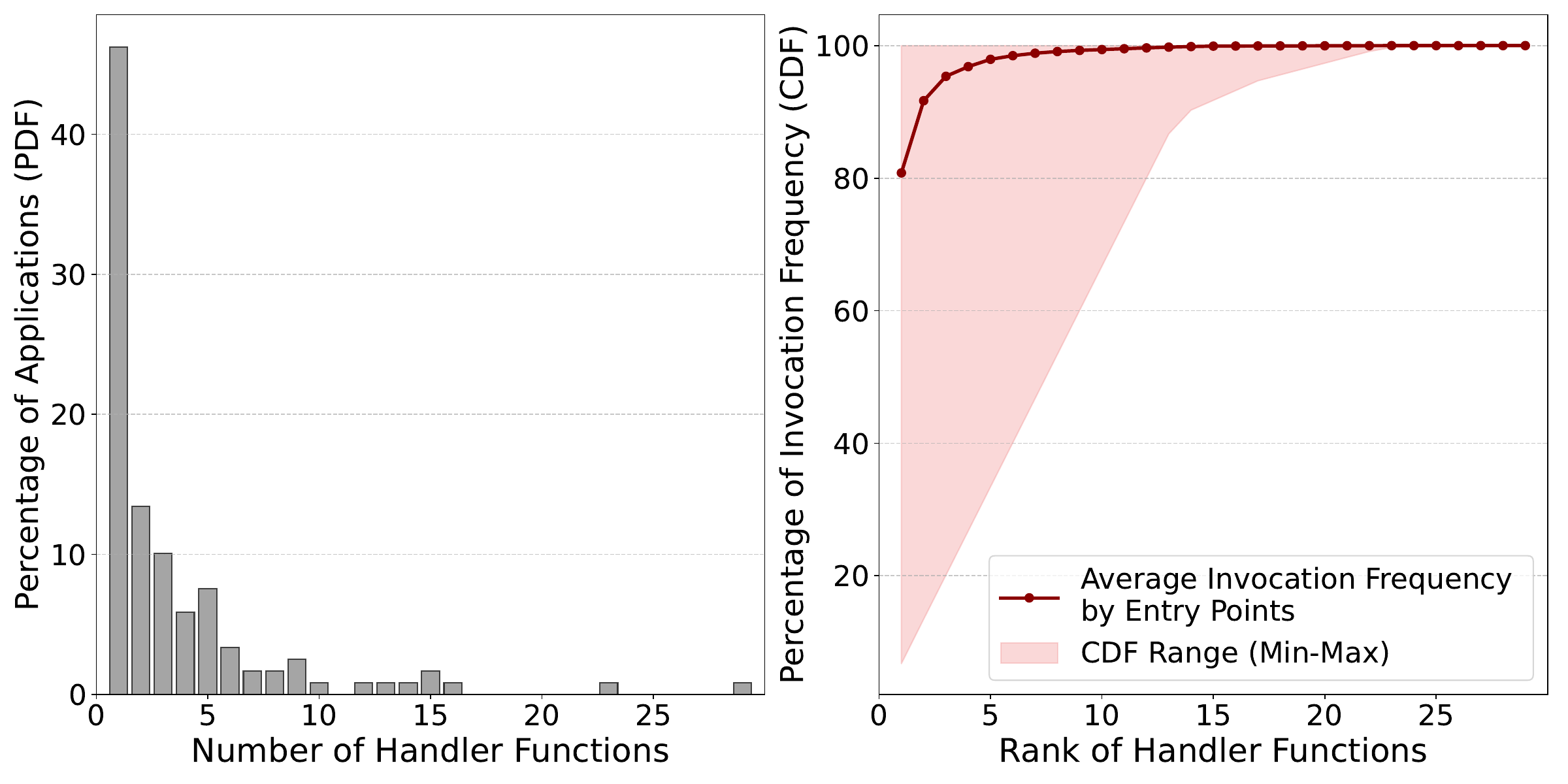}
    \caption{PDF plot of serverless applications by the number of handler functions and CDF of entry point invocation frequencies from production trace data.}
    \label{fig:realTrace}
\end{figure}

To further explore the potential impact of addressing this deficiency, we conducted additional studies using production traces~\cite{shahrad2020serverless}, 
which contains 119 serverless applications and their real-world requests. We summarize our findings as below: 

1) A majority of serverless functions have more than one entry function. Figure~\ref{fig:realTrace}(1) shows the probability density function (PDF) of serverless functions based on the number of entry functions, where 54\% of them include more than one entry function. The prevalence of multiple entry functions highlights the potential widespread existence of workload-dependent libraries.

2) Entry function invocation varies significantly, with only a small number of functions being used most of the time. Figure~\ref{fig:realTrace}{(2)} illustrates the cumulative distribution function (CDF) of entry function requests, showing that the top few handler functions account for over 80\% of cumulative invocations, while many others are rarely invoked. This skewed distribution suggests that libraries required by infrequently used functions are accessed less often, offering potential to optimize their loading to reduce latency further.

\begin{observation}
Serverless applications exhibit highly variable handler function usage, with a few entry points dominating invocations. This widespread variability highlights the potential prevalence of workload-dependent libraries and the significant opportunity to address the deficiencies of static-analysis-based library loading optimization. 
\end{observation}

%% file: sections/2.1-inefficiency-categorization.tex
\section{Design of \tool{}}
\label{design}

To effectively identify and optimize inefficient library usage in serverless applications, we design \tool{}, seamlessly integrating with serverless CI/CD pipelines. Figure~\ref{fig:architecture} depicts the overview of \tool{}.


\begin{figure}[!hbt]
    \centering
    \resizebox{0.5\textwidth}{!}{%
        \input{figures/workflow_overview}
    }
    \caption{\tool{} workflow overview.}
    \label{fig:architecture}
\end{figure}

At a high level, \tool{} consists of three main components: 1) a \textbf{dynamic profiler}, 2) an \textbf{automated code optimizer}, and 3) an \textbf{adaptive mechanism for evolving workloads}. The dynamic profiler monitors serverless application execution to collect runtime profiles of library usage patterns. It introduces a library utilization metric to guide code optimization decisions. The code optimizer then leverages this profiling data to identify unnecessary library initializations caused by global imports and replaces them with deferred imports, enabling reduced cold-start time. Finally, to ensure optimizations remain effective under changing execution patterns, the adaptive mechanism dynamically monitors workload variations and iteratively triggers profiling and optimization phases as application workloads evolve.

The design of \tool{} is guided by the need to address the following technical challenges (TC):  

\textit{TC-1: How to monitor library usage without incurring significant monitoring overhead:} Serverless applications are often latency-sensitive, especially those used in real-time scenarios such as APIs, event-driven processing, or user-facing applications, where even minor delays can degrade performance or user experience. Traditional monitoring techniques, such as binary or source code instrumentation, can introduce substantial overhead, making them unsuitable for such environments. 

\noindent\textbf{Solution:} 
To minimize monitoring overhead, \tool{} uses four strategies: (1) Lightweight profiling periodically samples and captures executed Python code during function execution, avoiding monitoring every instruction. (2) Aggregating samples across multiple invocations distributes the workload, lowering the impact on individual invocations. (3) Profiling data is collected locally and batch-transferred asynchronously to an external collector, minimizing network transmission overhead. (4) To further reduce the fine-grain profiling overhead, \tool{} employs a data-driven adaptive profiling mechanism that monitors workload variations over time and detects significant changes in function invocation patterns. When shifts in usage are identified, \tool{} triggers fine-grained code profiling and updates the code optimizer to maintain effectiveness. Details of the first three strategies are discussed in Section~\ref{dynamic_monitoring_of_library_initialization}, while Section~\ref{datadriven_adaptive_profiling} elaborates on the adaptive profiling mechanism.

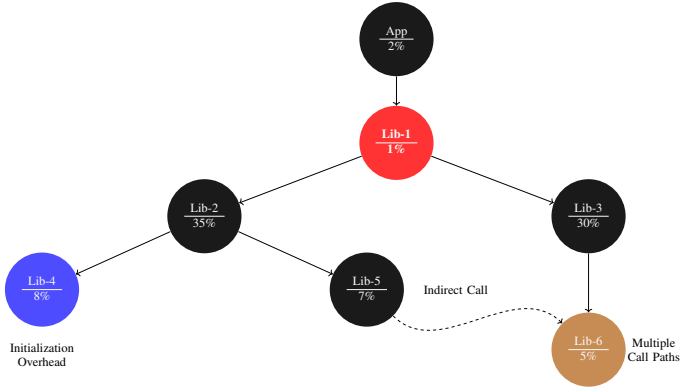
\begin{figure}[h!]
\centering
\resizebox{0.5\textwidth}{!}{%
\begin{tikzpicture}[node distance=3.5cm, auto]

\node (root) [circle, minimum size=2.5cm, text centered, fill=black!90, text=white, font=\large] {
  \begin{tabular}{c}
    App \\ \hline 2\%
  \end{tabular}
};

\node (orchestrator) [circle, minimum size=2.5cm, below of=root, text centered, fill=red!80, text=white, font=\large] {
  \begin{tabular}{c}
    \textbf{Lib-1} \\ \hline \textbf{1\%}
  \end{tabular}
};

\node (lib2) [circle, minimum size=2.5cm, below left of=orchestrator, xshift=-4cm, text centered, fill=black!90, text=white, font=\large] {
  \begin{tabular}{c}
    Lib-2 \\ \hline 35\%
  \end{tabular}
};
\node (lib3) [circle, minimum size=2.5cm, below right of=orchestrator, xshift=4cm, text centered, fill=black!90, text=white, font=\large] {
  \begin{tabular}{c}
    Lib-3 \\ \hline 30\%
  \end{tabular}
};

\node (lib4) [circle, minimum size=2.5cm, below left of=lib2, xshift=-3cm, text centered, fill=blue!70, text=white, font=\large] {
  \begin{tabular}{c}
    Lib-4 \\ \hline 8\%
  \end{tabular}
};
\node (lib5) [circle, minimum size=2.5cm, below right of=lib2, xshift=3cm, text centered, fill=black!90, text=white, font=\large] {
  \begin{tabular}{c}
    Lib-5 \\ \hline 7\%
  \end{tabular}
};
\node (lib6) [circle, minimum size=2.5cm, below of=lib3, yshift=-1cm, text centered, fill=brown!90, text=white, font=\large] {
  \begin{tabular}{c}
    Lib-6 \\ \hline 5\%
  \end{tabular}
};

\draw[->, thick] (root) -- (orchestrator);
\draw[->, thick] (orchestrator) -- (lib2);
\draw[->, thick] (orchestrator) -- (lib3);
\draw[->, thick] (lib2) -- (lib4);
\draw[->, thick] (lib2) -- (lib5);
\draw[->, thick] (lib3) -- (lib6);

\draw[->, dashed, thick] (lib5) to[out=315, in=135] (lib6);

\node [below of=lib4, yshift=1.3cm, font=\large, align=center] (note1) {Initialization \\ Overhead};
\node [above of=lib5, xshift=3cm, yshift=-3.5cm, font=\large, align=center] (note2) {Indirect Call};
\node [right of=lib6, xshift=-1.3cm, font=\large, align=center] (note3) {Multiple \\ Call Paths};

\end{tikzpicture}
}
\caption{Dependency graph with an entry point for a serverless function. The entry point (App) delegates to the orchestrator (Lib-1), which coordinates tasks among libraries with cascading dependencies. Lib-4 (blue) represents a library with all samples originating during initialization, demonstrating the challenge of attributing initialization vs. runtime usage. Dashed edges represent indirect calls, such as between Lib-5 and Lib-6. Lib-6 (Brown) is accessed via multiple call paths, illustrating the complexity of analyzing usage in multi-path scenarios.}
\label{fig:serverless_entrypoint_dependency_graph_with_annotations}
\end{figure}

\textit{TC-2: How to accurately identify libraries with significant initialization overhead that are rarely or never utilized during runtime:}  
Dynamically analyzing library usage is challenging due to several factors. First, libraries often invoke others, forming cascading dependencies where heavier callee libraries dominate profiling samples, while lightweight yet critical libraries like orchestrators are underrepresented. Profiling samples do not directly reflect library usage, making it challenging to accurately assess the contributions of each library. For instance, in Figure~\ref{fig:serverless_entrypoint_dependency_graph_with_annotations}, the orchestrator library (\texttt{Lib-1}) collects only 1\% of samples, despite coordinating all downstream tasks. Second, libraries can be accessed via multiple call paths, as demonstrated by \texttt{Lib-6}, which is invoked both directly and indirectly. This complexity complicates accurate usage attribution and can lead to incomplete profiling insights. Finally, profiling may conflate library access during initialization with runtime usage, as highlighted by \texttt{Lib-4}, where all samples are due to initialization rather than actual execution. Together, these issues make it challenging to assess the contributions of individual libraries from profiling data.

\noindent\textbf{Solution:} (1) To address cascading dependencies, \tool{} employs a Calling Context Tree (CCT), a data structure that represents the hierarchical caller-callee relationships during program execution. While CCTs are commonly used for visualizing call chains, \tool{} uniquely leverages this structure to accurately attribute profiling samples from callee libraries to their parent nodes along the call chain. By propagating samples upward through the CCT, \tool{} ensures these libraries are correctly attributed with the activity of their dependent libraries. (2) For handling multiple call paths, \tool{} uses the CCT to attribute profiling samples to their precise invocation paths, preserving the calling context for every library invocation. Unlike conventional profilers that aggregate metrics across paths, \tool{} distinguishes between different uses of the same library invoked through distinct paths. This level of granularity prevents conflating diverse usage patterns, enabling precise analysis of library behavior. (3)  To separate initialization samples from runtime execution samples, \tool{} identifies samples originating from \texttt{\_\_init\_\_} methods of the package. By isolating initialization phases, \tool{} avoids misleading metrics and ensures that optimizations focus on libraries with significant runtime impact. Section~\ref{dynamic_monitoring_of_library_initialization} discusses the solutions in detail.


%% file: figures/workflow_overview.tex
\tikzset{every picture/.style={line width=0.75pt}} 

\begin{tikzpicture}[x=0.75pt,y=0.75pt,yscale=-1,xscale=1]

\draw  [fill={rgb, 255:red, 255; green, 255; blue, 255 }  ,fill opacity=1 ] (98.33,29) -- (597.33,29) -- (597.33,273) -- (98.33,273) -- cycle ;
\draw  [fill={rgb, 255:red, 0; green, 0; blue, 0 }  ,fill opacity=1 ] (413.98,63.35) .. controls (413.98,56.22) and (419.76,50.44) .. (426.89,50.44) -- (563.75,50.44) .. controls (570.89,50.44) and (576.67,56.22) .. (576.67,63.35) -- (576.67,102.09) .. controls (576.67,109.22) and (570.89,115) .. (563.75,115) -- (426.89,115) .. controls (419.76,115) and (413.98,109.22) .. (413.98,102.09) -- cycle ;

\draw  [fill={rgb, 255:red, 0; green, 0; blue, 0 }  ,fill opacity=1 ] (272.98,208.35) .. controls (272.98,201.22) and (278.76,195.44) .. (285.89,195.44) -- (422.75,195.44) .. controls (429.89,195.44) and (435.67,201.22) .. (435.67,208.35) -- (435.67,247.09) .. controls (435.67,254.22) and (429.89,260) .. (422.75,260) -- (285.89,260) .. controls (278.76,260) and (272.98,254.22) .. (272.98,247.09) -- cycle ;

\draw  [fill={rgb, 255:red, 0; green, 0; blue, 0 }  ,fill opacity=1 ] (120.98,67.35) .. controls (120.98,60.22) and (126.76,54.44) .. (133.89,54.44) -- (270.75,54.44) .. controls (277.89,54.44) and (283.67,60.22) .. (283.67,67.35) -- (283.67,106.09) .. controls (283.67,113.22) and (277.89,119) .. (270.75,119) -- (133.89,119) .. controls (126.76,119) and (120.98,113.22) .. (120.98,106.09) -- cycle ;

\draw  [fill={rgb, 255:red, 140; green, 40; blue, 54 }  ,fill opacity=1 ] (343,50.5) -- (378,85.5) -- (343,120.5) -- (308,85.5) -- cycle ;

\draw    (378,85.5) -- (410.33,85.03) ;
\draw [shift={(412.33,85)}, rotate = 179.17] [color={rgb, 255:red, 0; green, 0; blue, 0 }  ][line width=0.75]    (10.93,-3.29) .. controls (6.95,-1.4) and (3.31,-0.3) .. (0,0) .. controls (3.31,0.3) and (6.95,1.4) .. (10.93,3.29)   ;
\draw    (282.33,85) -- (306,85.46) ;
\draw [shift={(308,85.5)}, rotate = 181.12] [color={rgb, 255:red, 0; green, 0; blue, 0 }  ][line width=0.75]    (10.93,-3.29) .. controls (6.95,-1.4) and (3.31,-0.3) .. (0,0) .. controls (3.31,0.3) and (6.95,1.4) .. (10.93,3.29)   ;
\draw    (343,120.5) .. controls (266.69,170.46) and (217.36,133.38) .. (199.86,120.16) ;
\draw [shift={(198.33,119)}, rotate = 36.87] [color={rgb, 255:red, 0; green, 0; blue, 0 }  ][line width=0.75]    (10.93,-3.29) .. controls (6.95,-1.4) and (3.31,-0.3) .. (0,0) .. controls (3.31,0.3) and (6.95,1.4) .. (10.93,3.29)   ;
\draw  [fill={rgb, 255:red, 140; green, 40; blue, 54 }  ,fill opacity=1 ] (503,142.5) -- (538,177.5) -- (503,212.5) -- (468,177.5) -- cycle ;

\draw    (502.33,117) -- (502.95,140.5) ;
\draw [shift={(503,142.5)}, rotate = 268.51] [color={rgb, 255:red, 0; green, 0; blue, 0 }  ][line width=0.75]    (10.93,-3.29) .. controls (6.95,-1.4) and (3.31,-0.3) .. (0,0) .. controls (3.31,0.3) and (6.95,1.4) .. (10.93,3.29)   ;
\draw    (468,177.5) .. controls (424.77,151.26) and (407.68,149.04) .. (357.85,190.72) ;
\draw [shift={(356.33,192)}, rotate = 319.86] [color={rgb, 255:red, 0; green, 0; blue, 0 }  ][line width=0.75]    (10.93,-3.29) .. controls (6.95,-1.4) and (3.31,-0.3) .. (0,0) .. controls (3.31,0.3) and (6.95,1.4) .. (10.93,3.29)   ;
\draw   (607.33,111) -- (634.42,111) -- (649,125.58) -- (649,172) -- (607.33,172) -- cycle ;
\draw   (634.42,111) -- (649,125.58) -- (634.42,125.58) -- cycle ;

\draw  [dash pattern={on 4.5pt off 4.5pt}]  (626.33,110) .. controls (624.36,84.39) and (616.57,74.3) .. (580.02,78.79) ;
\draw [shift={(578.33,79)}, rotate = 352.5] [color={rgb, 255:red, 0; green, 0; blue, 0 }  ][line width=0.75]    (10.93,-3.29) .. controls (6.95,-1.4) and (3.31,-0.3) .. (0,0) .. controls (3.31,0.3) and (6.95,1.4) .. (10.93,3.29)   ;
\draw  [dash pattern={on 4.5pt off 4.5pt}]  (628.33,172) .. controls (626.35,222.49) and (496.96,234.76) .. (440.04,231.12) ;
\draw [shift={(438.33,231)}, rotate = 4.09] [color={rgb, 255:red, 0; green, 0; blue, 0 }  ][line width=0.75]    (10.93,-3.29) .. controls (6.95,-1.4) and (3.31,-0.3) .. (0,0) .. controls (3.31,0.3) and (6.95,1.4) .. (10.93,3.29)   ;
\draw    (78.33,84) -- (117.33,84) ;
\draw [shift={(119.33,84)}, rotate = 180] [color={rgb, 255:red, 0; green, 0; blue, 0 }  ][line width=0.75]    (10.93,-3.29) .. controls (6.95,-1.4) and (3.31,-0.3) .. (0,0) .. controls (3.31,0.3) and (6.95,1.4) .. (10.93,3.29)   ;

\draw (10,84) node [anchor=north west][inner sep=0.75pt]   [align=left] {\begin{minipage}[lt]{44.57pt}\setlength\topsep{0pt}
\begin{center}
{\Large Function }\\{\Large Invocation}
\end{center}

\end{minipage}};
\draw (37,53) node[anchor=north west, inner sep=0.75pt, 
    text={rgb,255:red,208; green,2; blue,27}, opacity=1, align=left] {{\Huge $\lambda$}};

\draw (608,120) node [anchor=north west][inner sep=0.75pt]   [align=left] {\Large src};
\draw (405,138) node [anchor=north west][inner sep=0.75pt]   [align=left] {Yes};
\draw (255,148) node [anchor=north west][inner sep=0.75pt]   [align=left] {No};
\draw (379,67) node [anchor=north west][inner sep=0.75pt]   [align=left] {Yes};
\draw (116.42,247.63) node [anchor=north west][inner sep=0.75pt]   [align=left] {\Large \tool{}};
\draw (610,134) node [anchor=north west][inner sep=0.75pt]   [align=left] {$< / >$};
\draw (477,171) node [anchor=north west][inner sep=0.75pt]   [align=left] {\textcolor[rgb]{1,1,1}{{\small Optimize?}}};
\draw (322,77) node [anchor=north west][inner sep=0.75pt]   [align=left] {\textcolor[rgb]{1,1,1}{Profile?}};
\draw (138.98,59) node [anchor=north west][inner sep=0.75pt]  [color={rgb, 255:red, 255; green, 255; blue, 255 }  ,opacity=1 ] [align=left] {\begin{minipage}[lt]{88.15pt}\setlength\topsep{0pt}
\begin{center}
\large Adaptive Workload Monitor
\end{center}
\end{minipage}};
\draw (305.57,219) node [anchor=north west][inner sep=0.75pt]  [color={rgb, 255:red, 255; green, 255; blue, 255 }  ,opacity=1 ] [align=left] {\large Code Optimizer};
\draw (447,74) node [anchor=north west][inner sep=0.75pt]  [color={rgb, 255:red, 255; green, 255; blue, 255 }  ,opacity=1 ] [align=left] {\large Code Profiler};

\end{tikzpicture}

%% file: sections/5-Design-implementation.tex
\section{Methodology and Implementation}
\label{methodology}

To address the technical challenges outlined in the design, \tool{} employs a systematic methodology to profile serverless applications to identify inefficient library usage patterns and apply automated code optimizations. \tool{} further incorporates an adaptive mechanism to iteratively trigger profiling and optimization phases as workloads evolve.

\subsection{Dynamic Monitoring of Library Initialization and Usage}
\label{dynamic_monitoring_of_library_initialization}
To identify libraries that incur significant initialization time with low utilization, \tool{}'s dynamic monitoring is divided into two parts: (1) monitoring the time spent initializing dependent libraries and sub-modules, and (2) profiling library usage during runtime using statistical sampling. By combining these two phases, \tool{} derives two key metrics—initialization overhead and runtime utilization—for each library and its sub-modules, thus enabling precise identification of inefficient dependencies.


 
\subsubsection{Hierarchical Breakdown of Initialization Overhead}
\label{init_time_measurement}
\tool{} systematically identifies libraries and packages that contribute significantly to cold-start latency through a hierarchical breakdown of initialization overhead. This analysis begins by measuring the total initialization time of all libraries relative to the application end-to-end time to quantify their overall impact. To determine substantial overhead, \tool{} applies a predefined threshold, identifying cases where the total library initialization time exceeds 10\% of the application's end-to-end time. By selectively focusing on such applications, \tool{} avoids unnecessary profiling overhead. \tool{} then decomposes the initialization time into individual libraries and subsequently into their constituent packages and sub-packages.

\begin{figure}[h!]
\centering
\resizebox{0.5\textwidth}{!}{%
\begin{tikzpicture}[node distance=2cm, auto]

\node (app) [draw, rectangle, rounded corners, text centered, font=\small] {\begin{tabular}{c}
    Application \\
    \hline
    \textcircled{\small{1}} Init. to execution Ratio = 0.4 ($>$Threshold)
  \end{tabular}};
\node (libs) [draw, rectangle, rounded corners, below of=app, text centered, node distance=1.2cm, font=\small, fill=gray!80] {
  \begin{tabular}{c}
    Libraries \\
    \hline
    Init. overhead = 100\%
  \end{tabular}
};
\node (lib1) [draw, rectangle, rounded corners, below left of=libs, xshift=-3cm, text centered, node distance=1.7cm, font=\small, fill=gray!55] {
  \begin{tabular}{c}
    Library 1 \\
    \hline
    \textcircled{\small{2}} Init. overhead = 95\%
  \end{tabular}
};
\node (lib2) [draw, rectangle, rounded corners, below of=libs, xshift=3cm, text centered, node distance=1.2cm, font=\small, fill=gray!5] {
  \begin{tabular}{c}
    Library 2 \\
    \hline
    Init. overhead = 5\%
  \end{tabular}
};
\node (pkg1) [draw, rectangle, rounded corners, below of=lib1, text centered, node distance=1.2cm, font=\small, fill=gray!55] {
  \begin{tabular}{c}
    Library 1.pkg \\
    \hline
    Init. overhead = 95\%
  \end{tabular}
};
\node (pkg2) [draw, rectangle, rounded corners, below of=lib2, text centered, node distance=1.2cm, font=\small, fill=gray!5] {
  \begin{tabular}{c}
    Library 2.pkg \\
    \hline
    Init. overhead = 5\%
  \end{tabular}
};
\node (subpkg1) [draw, rectangle, rounded corners, below left of=pkg1, xshift=-1.5cm, text centered, node distance=2cm, font=\small, fill=gray!55] {
  \begin{tabular}{c}
    Library 1.pkg.subpkg1 \\
    \hline
    \textcircled{\small{3}} Init. overhead = 85\%
  \end{tabular}
};
\node (subpkg2) [draw, rectangle, rounded corners, below right of=pkg1, xshift=1.5cm, text centered, node distance=2cm, font=\small, fill=gray!5] {
  \begin{tabular}{c}
    Library 1.pkg.subpkg2 \\
    \hline
    Init. overhead = 10\%
  \end{tabular}
};

\draw[->] (app) -- (libs);
\draw[->] (libs) -- (lib1);
\draw[->] (libs) -- (lib2);
\draw[->] (lib1) -- (pkg1);
\draw[->] (lib2) -- (pkg2);
\draw[->] (pkg1) -- (subpkg1);
\draw[->] (pkg1) -- (subpkg2);

\end{tikzpicture}
}

\caption{Hierarchical analysis policy with initialization overhead. Darker nodes represent higher overhead leading to increased cold-start time.}
\label{fig:hierarchical_analysis}
\end{figure}
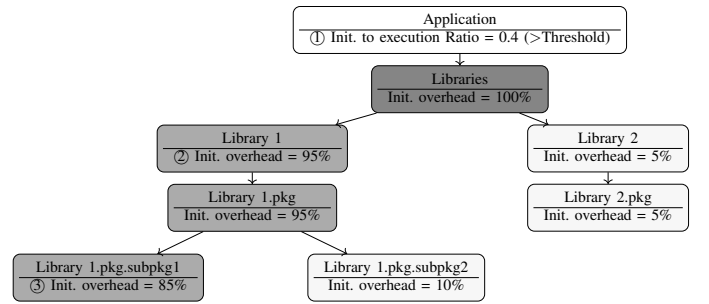

\tool{} measures three types of initialization times, as shown in figure~
\ref{fig:hierarchical_analysis}: (1) the total initialization overhead across all loaded libraries, (2) the cumulative initialization time for each library by summing up the time taken by its modules, and (3) the initialization time for packages and sub-packages. The accumulated initialization overhead \( T_{\text{total\_initialization}} \) for \( L \) libraries is calculated as:
\begin{equation}
T_{\text{total\_initialization}} = \sum_{k=1}^{L} T_{\text{library}_k}.
\end{equation}

The library-level initialization time \( T_{\text{library}} \), derived from \( N \) modules, is:
\begin{equation}
T_{\text{library}} = \sum_{i=1}^{N} T_i.
\end{equation}

Finally, the initialization time \( T_{\text{package}} \) of a package with \( M \) modules, where \( T_j \) is the initialization time of the \( j \)-th module, is:
\begin{equation}
T_{\text{package}} = \sum_{j=1}^{M} T_j.
\end{equation}
This hierarchical decomposition provides detailed insights into initialization delays, thus guiding effective optimization efforts.

\subsubsection{Sampling-Based Call-path Profiling}
\label{sampling_based_callpath_profiling}
To monitor the library usage of a serverless application, we employ sampling-based call-path profiling. \tool{} sets up a timer with a configurable sampling frequency and registers a signal handler. When the timer expires, it triggers the signal handler, enabling \tool{} to capture the application's current state at that moment. \tool{} gathers the following data in the signal handler: (1) the source code line number, (2) the function name, (3) file path and (4) the call path leading to the function call. At the implementation level, \tool{} utilizes Python's native \texttt{traceback} module to extract the frames from the Python call stack and construct the call path.

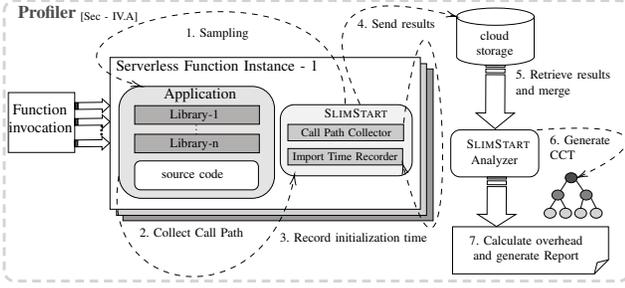
\begin{figure}[!t]
\setlength{\belowcaptionskip}{-15pt}
    \centering
    \resizebox{0.5\textwidth}{!}{%
        \input{figures/profiler_design_overview}
    }
    \caption{\tool{} profiler architecture overview.}
    \label{fig:profiler_design}
\end{figure}


\noindent{\textbf{Calling context tree (CCT):}}
\tool{} accumulates the call paths to construct a Calling Context Tree (CCT)~\cite{ammons1997exploiting}, a data structure that captures the function calling context, as shown in figure~\ref{fig:profiler_design}. In the CCT, each node represents a function call, and the edges represent the caller-callee relationship. The root of the tree represents the entry point of the serverless function. By preserving unique call paths for each invocation, the CCT distinguishes between different calling contexts of the same function. This ensures that usage attribution remains accurate even when functions or libraries are invoked from multiple locations in the code.

To support precise analysis, \tool{} augments the CCT with \textbf{sample counts} collected during statistical sampling. These counts measure the frequency of function executions under the observed workload. The sample counts at each node are escalated up the tree, propagating usage information toward the root. This escalation ensures that libraries invoked indirectly through cascading dependencies are appropriately attributed to their respective callers, thus eliminating the attribution challenge discussed in section~\ref{design}. The augmented CCT allows \tool{} to pinpoint inefficient invocations and trace their origins, finally guiding code optimizations.


\noindent{\textbf{Associating the Call-paths to Libraries:}}
To associate function invocations to libraries, \tool{} maps the nodes of the CCT to the corresponding library modules. This mapping is achieved by leveraging file paths associated with each function call in the CCT. Python libraries, organized hierarchically into packages, sub-packages, and modules, are accurately labeled at the node level. By propagating sample counts and associating call paths with libraries, \tool{} can attribute both direct and indirect library usage to their respective calling contexts.

To further refine this attribution, \tool{} distinguishes samples originating from library initialization from actual runtime usage. It achieves this by traversing the call chain of each sampled event to detect any \texttt{\_\_init\_\_} calls, filtering out samples associated with initialization.

\noindent{\textbf{Measuring the Library Utilization:}} Finally, \tool{} calculates a utilization metric for each library. This metric is defined as:

\begin{equation}
U(L) = \frac{\sum_{f \in L} S(f)}{\sum_{f \in F} S(f)}
\end{equation}

where \( U(L) \) is the utilization of library \( L \), \( S(f) \) is the sample count for function \( f \), \( f \in L \) represents all functions in library \( L \), and \( f \in F \) represents all functions in the application. This metric quantifies the execution frequency of functions from a specific library relative to the entire application. This enables the identification of both heavily used and underutilized libraries to prioritize optimization efforts.

\noindent{\textbf{Detecting inefficient library usage:}} \tool{} combines package initialization time and the library utilization metric to identify inefficient libraries. Libraries are ranked by initialization latency, and those with significant overhead but no usage samples are flagged as unused. Additionally, libraries with high initialization latency but low utilization, determined using a threshold of 2\% of the collected samples, are identified as infrequently used. These insights help pinpoint libraries that contribute to cold-start overhead without delivering proportional runtime benefits.

\subsection{Code Optimization}
Manually identifying and optimizing library imports in serverless applications is a time-consuming and error-prone task, especially in applications with complex code structures and dependencies. However, automating code transformations introduces its own challenges, particularly ensuring that the transformations preserve functional correctness. Therefore, an automated solution must carefully analyze the code structure to detect global imports and introduce changes only where they are safe and effective.

To address this, \tool{} implements an automatic code optimizer that leverages Python's Abstract Syntax Tree (AST) to analyze the application code. For each inefficient library flagged, the optimizer detects global imports and replaces them with deferred imports at their first usage points. By commenting out global imports and introducing deferred imports seamlessly across the codebase, \tool{} ensures functional correctness while adhering to coding standards.


\subsection{Adaptive Mechanism for Evolving Workloads}
\label{datadriven_adaptive_profiling}
After code optimization is performed, iterative profiling is necessary to ensure that further inefficiencies are identified as invocation patterns evolve. To minimize profiling overhead, \tool{} employs a data-driven adaptive profiling mechanism, which dynamically decides when to trigger profiling based on workload behavior.

\tool{} continuously tracks the invocation frequencies of handler functions. Let $f_i$ represent an entry handler function, and $p_i(t)$ denote the probability of $f_i$ being invoked at time $t$. The probability is calculated as:
\begin{equation}
p_i(t) = \frac{N_i(t)}{\sum_{j=1}^n N_j(t)}
\end{equation}
where $N_i(t)$ is the number of invocations of $f_i$ within a time window $[t-\Delta t, t]$, and $n$ is the total number of handler functions.

To detect significant changes in invocation patterns, \tool{} computes the difference in probabilities $\Delta p_i(t)$ over successive time intervals:
\begin{equation}
\Delta p_i(t) = p_i(t) - p_i(t - \Delta t).
\end{equation}
Profiling is triggered if the aggregate change across all handler functions exceeds a predefined threshold $\epsilon$:
\begin{equation}
\sum_{i=1}^n |\Delta p_i(t)| > \epsilon.
\end{equation}

The parameters $\Delta t$ and $\epsilon$ can be dynamically adjusted based on observed workload characteristics. Smaller $\epsilon$ values increase sensitivity to changes, while larger $\Delta t$ values provide smoother estimates of invocation patterns. By focusing profiling efforts on periods of significant workload changes, this mechanism reduces unnecessary profiling overhead while maintaining accuracy in identifying inefficiencies.

\subsection{Implementation}
\tool{} is implemented as a Python module, designed to seamlessly integrate into serverless applications to monitor library usage patterns. At runtime, \tool{} performs call-path sampling to monitor library initialization time and track function executions with minimal overhead. To reduce the per-invocation monitoring overhead further, \tool{} aggregates sampled data across multiple function invocations, distributing the profiling workload and reducing the impact on individual executions. Additionally, profiling data is collected locally and batch-transferred asynchronously to external storage services, such as AWS DynamoDB or S3, to minimize network transmission overhead. Once the data is collected, \tool{} runs a background service to perform the analysis, constructing a Calling Context Tree (CCT) to preserve the calling context and calculate library utilization metrics. These insights guide the automated code optimizer, which identifies inefficient global imports and transforms them into deferred imports to reduce initialization overhead and improve cold-start performance.


%% file: figures/profiler_design_overview.tex
  
\tikzset {_1ikv98ihb/.code = {\pgfsetadditionalshadetransform{ \pgftransformshift{\pgfpoint{89.1 bp } { -108.9 bp }  }  \pgftransformscale{1.32 }  }}}
\pgfdeclareradialshading{_0bmddzar1}{\pgfpoint{-72bp}{88bp}}{rgb(0bp)=(1,1,1);
rgb(0bp)=(1,1,1);
rgb(25bp)=(0,0,0);
rgb(400bp)=(0,0,0)}
\tikzset{every picture/.style={line width=0.75pt}} 

\begin{tikzpicture}[x=0.75pt,y=0.75pt,yscale=-1,xscale=1]
\clip (0, 300) rectangle (690, 650);

\draw  [fill={rgb, 255:red, 155; green, 155; blue, 155 }  ,fill opacity=1 ] (126.16,402.52) -- (450.56,402.52) -- (450.56,561.33) -- (126.16,561.33) -- cycle ;
\draw  [fill={rgb, 255:red, 214; green, 210; blue, 211 }  ,fill opacity=1 ] (119.42,396.96) -- (443.67,396.96) -- (443.67,555.78) -- (119.42,555.78) -- cycle ;
\draw  [fill={rgb, 255:red, 255; green, 255; blue, 255 }  ,fill opacity=1 ] (112.66,390.5) -- (437.22,390.5) -- (437.22,549.33) -- (112.66,549.33) -- cycle ;
\draw  [fill={rgb, 255:red, 179; green, 177; blue, 177 }  ,fill opacity=0.34 ] (121.98,444.19) .. controls (121.98,431.26) and (132.47,420.77) .. (145.4,420.77) -- (261.25,420.77) .. controls (274.18,420.77) and (284.67,431.26) .. (284.67,444.19) -- (284.67,514.45) .. controls (284.67,527.38) and (274.18,537.87) .. (261.25,537.87) -- (145.4,537.87) .. controls (132.47,537.87) and (121.98,527.38) .. (121.98,514.45) -- cycle ;
\draw  [fill={rgb, 255:red, 126; green, 121; blue, 121 }  ,fill opacity=0.49 ] (138.47,440.95) -- (268.86,440.95) -- (268.86,460.33) -- (138.47,460.33) -- cycle ;

\draw  [fill={rgb, 255:red, 126; green, 121; blue, 121 }  ,fill opacity=0.49 ] (138.41,470.87) -- (268.8,470.87) -- (268.8,490.26) -- (138.41,490.26) -- cycle ;

\draw  [dash pattern={on 0.84pt off 2.51pt}]  (202.82,461.01) -- (202.82,471.21) ;

\draw  [fill={rgb, 255:red, 255; green, 255; blue, 255 }  ,fill opacity=1 ] (138.41,504.81) .. controls (138.41,501.47) and (141.12,498.76) .. (144.47,498.76) -- (261.81,498.76) .. controls (265.16,498.76) and (267.87,501.47) .. (267.87,504.81) -- (267.87,522.97) .. controls (267.87,526.32) and (265.16,529.03) .. (261.81,529.03) -- (144.47,529.03) .. controls (141.12,529.03) and (138.41,526.32) .. (138.41,522.97) -- cycle ;

\draw  [fill={rgb, 255:red, 212; green, 212; blue, 212 }  ,fill opacity=0.29 ] (290.74,454.83) .. controls (290.74,446.61) and (297.4,439.94) .. (305.62,439.94) -- (413.78,439.94) .. controls (422,439.94) and (428.67,446.61) .. (428.67,454.83) -- (428.67,499.48) .. controls (428.67,507.7) and (422,514.37) .. (413.78,514.37) -- (305.62,514.37) .. controls (297.4,514.37) and (290.74,507.7) .. (290.74,499.48) -- cycle ;
\draw  [fill={rgb, 255:red, 155; green, 155; blue, 155 }  ,fill opacity=0.45 ] (298.67,460.56) -- (419.67,460.56) -- (419.67,477.7) -- (298.67,477.7) -- cycle ;

\draw  [fill={rgb, 255:red, 155; green, 155; blue, 155 }  ,fill opacity=0.45 ] (298.67,485.84) -- (418.67,485.84) -- (418.67,502.98) -- (298.67,502.98) -- cycle ;

\draw   (6.33,427.23) -- (75.76,427.23) -- (75.76,489.92) -- (6.33,489.92) -- cycle ;

\draw   (78.9,476.89) -- (104.18,476.59) -- (104.16,473.03) -- (112.14,480.06) -- (104.24,487.28) -- (104.22,483.71) -- (78.93,484.01) -- cycle ;\draw   (75.33,476.93) -- (76.05,476.92) -- (76.08,484.05) -- (75.37,484.06) -- cycle ;\draw   (76.76,476.91) -- (78.18,476.9) -- (78.22,484.02) -- (76.8,484.04) -- cycle ;
\draw   (79.09,454.48) -- (104.52,454.17) -- (104.51,450.61) -- (112.52,457.64) -- (104.58,464.86) -- (104.56,461.3) -- (79.13,461.6) -- cycle ;\draw   (75.53,454.52) -- (76.24,454.51) -- (76.28,461.64) -- (75.57,461.64) -- cycle ;\draw   (76.96,454.5) -- (78.38,454.48) -- (78.42,461.61) -- (77,461.63) -- cycle ;
\draw   (79.3,437.7) -- (104.49,437.4) -- (104.47,433.84) -- (112.42,440.87) -- (104.55,448.09) -- (104.53,444.52) -- (79.34,444.82) -- cycle ;\draw   (75.74,437.74) -- (76.45,437.73) -- (76.49,444.86) -- (75.77,444.86) -- cycle ;\draw   (77.16,437.72) -- (78.59,437.71) -- (78.62,444.83) -- (77.2,444.85) -- cycle ;

\draw  [dash pattern={on 4.5pt off 4.5pt}]  (121.98,514.45) .. controls (98,646.17) and (282,664.17) .. (305.62,513.03) ;
\draw [shift={(305.62,513.03)}, rotate = 98.88] [color={rgb, 255:red, 0; green, 0; blue, 0 }  ][line width=0.75]    (10.93,-3.29) .. controls (6.95,-1.4) and (3.31,-0.3) .. (0,0) .. controls (3.31,0.3) and (6.95,1.4) .. (10.93,3.29)   ;
\draw  [dash pattern={on 4.5pt off 4.5pt}]  (559.09,466.33) .. controls (678.95,457.16) and (673.39,483.07) .. (602.22,515.91) ;
\draw [shift={(601.15,516.41)}, rotate = 335.39] [color={rgb, 255:red, 0; green, 0; blue, 0 }  ][line width=0.75]    (10.93,-3.29) .. controls (6.95,-1.4) and (3.31,-0.3) .. (0,0) .. controls (3.31,0.3) and (6.95,1.4) .. (10.93,3.29)   ;
\draw  [dash pattern={on 4.5pt off 4.5pt}]  (358.56,438.96) .. controls (360,263.5) and (-34,371.5) .. (145.4,420.77) ;
\draw [shift={(145.4,420.77)}, rotate = 195.36] [color={rgb, 255:red, 0; green, 0; blue, 0 }  ][line width=0.75]    (10.93,-3.29) .. controls (6.95,-1.4) and (3.31,-0.3) .. (0,0) .. controls (3.31,0.3) and (6.95,1.4) .. (10.93,3.29)   ;
\draw  [fill={rgb, 255:red, 74; green, 74; blue, 74 }  ,fill opacity=1 ] (588.78,516.41) .. controls (588.78,513.54) and (591.55,511.22) .. (594.96,511.22) .. controls (598.38,511.22) and (601.15,513.54) .. (601.15,516.41) .. controls (601.15,519.27) and (598.38,521.59) .. (594.96,521.59) .. controls (591.55,521.59) and (588.78,519.27) .. (588.78,516.41) -- cycle ;
\draw  [fill={rgb, 255:red, 128; green, 128; blue, 128 }  ,fill opacity=1 ] (574.81,534.46) .. controls (574.81,531.59) and (577.58,529.27) .. (581,529.27) .. controls (584.41,529.27) and (587.18,531.59) .. (587.18,534.46) .. controls (587.18,537.32) and (584.41,539.64) .. (581,539.64) .. controls (577.58,539.64) and (574.81,537.32) .. (574.81,534.46) -- cycle ;
\draw  [fill={rgb, 255:red, 214; green, 210; blue, 211 }  ,fill opacity=1 ] (566.67,553.48) .. controls (566.67,550.62) and (569.43,548.3) .. (572.85,548.3) .. controls (576.26,548.3) and (579.03,550.62) .. (579.03,553.48) .. controls (579.03,556.35) and (576.26,558.67) .. (572.85,558.67) .. controls (569.43,558.67) and (566.67,556.35) .. (566.67,553.48) -- cycle ;
\draw  [fill={rgb, 255:red, 214; green, 210; blue, 211 }  ,fill opacity=1 ] (582.96,553.48) .. controls (582.96,550.62) and (585.73,548.3) .. (589.14,548.3) .. controls (592.56,548.3) and (595.33,550.62) .. (595.33,553.48) .. controls (595.33,556.35) and (592.56,558.67) .. (589.14,558.67) .. controls (585.73,558.67) and (582.96,556.35) .. (582.96,553.48) -- cycle ;
\draw    (578.45,539.64) -- (572.85,548.3) ;
\draw    (583.98,538.91) -- (589.14,548.3) ;

\draw  [fill={rgb, 255:red, 128; green, 128; blue, 128 }  ,fill opacity=1 ] (604.49,534.46) .. controls (604.49,531.59) and (607.26,529.27) .. (610.68,529.27) .. controls (614.09,529.27) and (616.86,531.59) .. (616.86,534.46) .. controls (616.86,537.32) and (614.09,539.64) .. (610.68,539.64) .. controls (607.26,539.64) and (604.49,537.32) .. (604.49,534.46) -- cycle ;
\draw  [fill={rgb, 255:red, 214; green, 210; blue, 211 }  ,fill opacity=1 ] (598.67,553.48) .. controls (598.67,550.62) and (601.44,548.3) .. (604.86,548.3) .. controls (608.27,548.3) and (611.04,550.62) .. (611.04,553.48) .. controls (611.04,556.35) and (608.27,558.67) .. (604.86,558.67) .. controls (601.44,558.67) and (598.67,556.35) .. (598.67,553.48) -- cycle ;
\draw  [fill={rgb, 255:red, 214; green, 210; blue, 211 }  ,fill opacity=1 ] (614.97,553.48) .. controls (614.97,550.62) and (617.74,548.3) .. (621.15,548.3) .. controls (624.57,548.3) and (627.33,550.62) .. (627.33,553.48) .. controls (627.33,556.35) and (624.57,558.67) .. (621.15,558.67) .. controls (617.74,558.67) and (614.97,556.35) .. (614.97,553.48) -- cycle ;
\draw    (607.26,539.15) -- (604.86,548.3) ;
\draw    (615.69,537.69) -- (621.15,548.3) ;

\draw [shading=_0bmddzar1,_1ikv98ihb]   (590.38,519.88) -- (583.98,529.88) ;
\draw    (600.27,519.39) -- (607.55,529.88) ;

\draw  [dash pattern={on 4.5pt off 4.5pt}]  (419.67,459.23) .. controls (459,154.83) and (488,764.17) .. (418.67,501.64) ;
\draw [shift={(418.67,501.64)}, rotate = 75.21] [color={rgb, 255:red, 0; green, 0; blue, 0 }  ][line width=0.75]    (10.93,-3.29) .. controls (6.95,-1.4) and (3.31,-0.3) .. (0,0) .. controls (3.31,0.3) and (6.95,1.4) .. (10.93,3.29)   ;
\draw   (556.22,347.78) -- (556.22,389.78) .. controls (556.22,394.75) and (538.09,398.78) .. (515.72,398.78) .. controls (493.35,398.78) and (475.22,394.75) .. (475.22,389.78) -- (475.22,347.78) .. controls (475.22,342.81) and (493.35,338.78) .. (515.72,338.78) .. controls (538.09,338.78) and (556.22,342.81) .. (556.22,347.78) .. controls (556.22,352.75) and (538.09,356.78) .. (515.72,356.78) .. controls (493.35,356.78) and (475.22,352.75) .. (475.22,347.78) ;

\draw   (469,476.17) .. controls (469,470.74) and (473.4,466.33) .. (478.83,466.33) -- (551.17,466.33) .. controls (556.6,466.33) and (561,470.74) .. (561,476.17) -- (561,505.67) .. controls (561,511.1) and (556.6,515.5) .. (551.17,515.5) -- (478.83,515.5) .. controls (473.4,515.5) and (469,511.1) .. (469,505.67) -- cycle ;

\draw   (525.69,408.55) -- (525.5,448.71) -- (535.5,448.76) -- (515.42,465.12) -- (495.5,448.57) -- (505.5,448.62) -- (505.69,408.45) -- cycle ;\draw   (525.74,398.55) -- (525.73,400.55) -- (505.73,400.45) -- (505.74,398.45) -- cycle ;\draw   (525.72,402.55) -- (525.7,406.55) -- (505.7,406.45) -- (505.72,402.45) -- cycle ;
\draw   (619.2,617) -- (470.89,617) -- (470.89,567.67) -- (634,567.67) -- (634,602.2) -- cycle -- (619.2,617) ; \draw   (634,602.2) -- (622.16,605.16) -- (619.2,617) ;
\draw   (527.12,525.71) -- (526.98,554.73) -- (536.98,554.78) -- (516.92,567.49) -- (496.98,554.59) -- (506.98,554.64) -- (507.12,525.62) -- cycle ;\draw   (527.16,515.71) -- (527.16,517.71) -- (507.16,517.62) -- (507.17,515.62) -- cycle ;\draw   (527.15,519.71) -- (527.13,523.71) -- (507.13,523.62) -- (507.15,519.62) -- cycle ;
\draw  [dash pattern={on 4.5pt off 4.5pt}]  (374,439.5) .. controls (347,378.5) and (328,308.83) .. (475.22,347.78) ;
\draw [shift={(475.22,347.78)}, rotate = 194.82] [color={rgb, 255:red, 0; green, 0; blue, 0 }  ][line width=0.75]    (10.93,-3.29) .. controls (6.95,-1.4) and (3.31,-0.3) .. (0,0) .. controls (3.31,0.3) and (6.95,1.4) .. (10.93,3.29)   ;
\draw  [color={rgb, 255:red, 214; green, 210; blue, 211 }  ,draw opacity=1 ][fill={rgb, 255:red, 252; green, 3; blue, 3 }  ,fill opacity=0 ][dash pattern={on 6.75pt off 4.5pt}][line width=2.25]  (2,342) .. controls (2,335.92) and (6.92,331) .. (13,331) -- (645,331) .. controls (651.08,331) and (656,335.92) .. (656,342) -- (656,613.5) .. controls (656,619.58) and (651.08,624.5) .. (645,624.5) -- (13,624.5) .. controls (6.92,624.5) and (2,619.58) .. (2,613.5) -- cycle ;

\draw (190.33,358) node [anchor=north west][inner sep=0.75pt]   [align=left] {{\normalsize 1. Sampling}};
\draw (142.33,568) node [anchor=north west][inner sep=0.75pt]   [align=left] {{\normalsize 2. Collect Call Path}};
\draw (289,572.33) node [anchor=north west][inner sep=0.75pt]   [align=left] {{\normalsize 3. Record initialization time}};
\draw (370.33,349.33) node [anchor=north west][inner sep=0.75pt]   [align=left] {{\normalsize 4. Send results}};
\draw (117.57,391.68) node [anchor=north west][inner sep=0.75pt]   [align=left] {{\large Serverless Function Instance - 1}};
\draw (535.74,402.54) node [anchor=north west][inner sep=0.75pt]   [align=left] {5. Retrieve results \\and merge};
\draw (571.56,470.8) node [anchor=north west][inner sep=0.75pt]   [align=left] {6. Generate \\CCT};
\draw (480.22,575.67) node [anchor=north west][inner sep=0.75pt]   [align=left] {\begin{minipage}[lt]{99.12pt}\setlength\topsep{0pt}
\normalsize
\begin{center}
7. Calculate overhead\\and generate Report
\end{center}

\end{minipage}};
\draw (107.5,342.42) node   [align=left] {\begin{minipage}[lt]{136.68pt}\setlength\topsep{0pt}
\textbf{\textcolor[rgb]{0.29,0.29,0.29}{{\Large Profiler}}}
\end{minipage}};
\draw (335.22,442.92) node [anchor=north west][inner sep=0.75pt]   [align=left] {{\normalsize \tool{}}};
\draw (80,341.33) node [anchor=north west][inner sep=0.75pt]  [font=\small] [align=left] {{\small [Sec - IV.A]}};
\draw (483.94,468.42) node [anchor=north west][inner sep=0.75pt]   [align=left] {\begin{minipage}[lt]{44.09pt}\setlength\topsep{0pt}
\begin{center}
\tool{}\\Analyzer
\end{center}

\end{minipage}};
\draw (489,361) node [anchor=north west][inner sep=0.75pt]   [align=left] {\begin{minipage}[lt]{36.75pt}\setlength\topsep{0pt}
\begin{center}
cloud\\storage
\end{center}

\end{minipage}};
\draw (107.05,459.14) node [anchor=north west][inner sep=0.75pt]  [rotate=-90.81] [align=left] {\textbf{{\large ...}}};
\draw (5.32,437.41) node [anchor=north west][inner sep=0.75pt]   [align=left] {\begin{minipage}[lt]{48.65pt}\setlength\topsep{0pt}
\large
\begin{center}
Function \\invocation
\end{center}

\end{minipage}};
\draw (302.2,488.5) node [anchor=north west][inner sep=0.75pt]   [align=left] {{\small Import Time Recorder}};
\draw (311.05,462.97) node [anchor=north west][inner sep=0.75pt]   [align=left] {{\small Call Path Collector}};
\draw (167.57,421.24) node [anchor=north west][inner sep=0.75pt]   [align=left] {{\large Application}};
\draw (165.08,505.72) node [anchor=north west][inner sep=0.75pt]   [align=left] {source code};
\draw (173.93,473.39) node [anchor=north west][inner sep=0.75pt]   [align=left] {Library-n};
\draw (173.99,443.47) node [anchor=north west][inner sep=0.75pt]   [align=left] {Library-1};

\end{tikzpicture}

%% file: sections/6-Evaluation.tex
\section{Evaluation}
\label{eval}
In this section, we assess \tool{} focusing on answering the following evaluation questions:
\begin{itemize}[leftmargin=*]
    \item \textbf{Q1 (Speedup):} How much performance improvement can be expected through \tool{}-guided optimization?
    \item \textbf{Q2 (Comparison):} How does \tool{} perform compared to existing tools like FaaSLight in identifying inefficient library usage and reducing cold-start overhead?
    \item \textbf{Q3 (Overhead):} How much overhead \tool{}-Profiler imposes in its default setting?
\end{itemize}

\begin{table*}
\setlength{\belowcaptionskip}{-5pt}
\fontsize{6.5pt}{8.5pt}\selectfont
\begin{tabular}{@{}cccccccccc@{}}
\hline
\multicolumn{6}{c}{\textbf{Program Information}} & \multicolumn{4}{c}{\textbf{Speedup}} \\ \hline
\textbf{Applications} & \textbf{Library} & \textbf{Type} & \textbf{\begin{tabular}[c]{@{}c@{}}\# of \\ libs\end{tabular}} & \textbf{\begin{tabular}[c]{@{}c@{}}\# of \\ modules\end{tabular}} & \textbf{\begin{tabular}[c]{@{}c@{}}Avg. \\ Depth\end{tabular}} & \textbf{\begin{tabular}[c]{@{}c@{}}Initialization\\Speedup \\(times)\end{tabular}} & \textbf{\begin{tabular}[c]{@{}c@{}}Execution \\ Speedup \\ (times)\end{tabular}} & \textbf{\begin{tabular}[c]{@{}c@{}}99\textsuperscript{th} Percentile \\ Initialization \\Speedup\end{tabular}} & \textbf{\begin{tabular}[c]{@{}c@{}}99\textsuperscript{th} Percentile \\ End-to-end \\Speedup\end{tabular}} \\ \hline 
\multicolumn{10}{c}{\textbf{RainbowCake Applications}} \\ \hline 
Dna-visualisation (R-DV) & NumPy & Scientific Computing & 2 & 242 & 4.75 & 2.30$\times$ & 2.26$\times$ & 2.03$\times$ & 1.99$\times$ \\
Graph-bfs (R-GB) & igraph & Graph Processing & 1 & 86 & 3.74 & 1.71$\times$ & 1.66$\times$ & 1.55$\times$ & 1.54$\times$ \\
Graph-mst (R-GM) & igraph & Graph Processing & 1 & 86 & 3.74 & 1.74$\times$ & 1.70$\times$ & 1.67$\times$ & 1.64$\times$ \\
Graph-pagerank (R-GPR) & igraph & Graph Processing & 1 & 86 & 3.74 & 1.70$\times$ & 1.62$\times$ & 1.69$\times$ & 1.64$\times$ \\
Sentiment-analysis (R-SA) & nltk, TextBlob & Natural Language Processing & 4 & 265 & 5.13 & 1.35$\times$ & 1.33$\times$ & 1.37$\times$ & 1.34$\times$ \\\hline
\multicolumn{10}{c}{\textbf{FaaSLight Applications}} \\ \hline
Price-ml-predict (FL-PMP) & SciPy & Machine Learning & 3 & 832 & 7.98 & 1.31$\times$ & 1.30$\times$ & 1.37$\times$ & 1.36$\times$ \\
Skimage-numpy (FL-SN) & SciPy & Image Processing & 14 & 656 & 5.32 & 1.41$\times$ & 1.36$\times$ & 1.41$\times$ & 1.37$\times$ \\
Predict-wine-ml (FL-PWM) & pandas & Machine Learning & 6 & 1385 & 7.57 & 1.76$\times$ & 1.68$\times$ & 1.59$\times$ & 1.52$\times$ \\
Train-wine-ml (FL-TWM) & pandas & Machine Learning & 6 & 1385 & 7.57 & 1.79$\times$ & 1.50$\times$ & 1.72$\times$ & 1.46$\times$ \\
Sentiment-analysis (FL-SA) & pandas, SciPy & Natural Language Processing & 6 & 1081 & 6.8 & 2.01$\times$ & 2.01$\times$ & 2.15$\times$ & 2.15$\times$ \\\hline
\multicolumn{10}{c}{\textbf{FaaS Workbench Applications}} \\ \hline
Chameleon (FWB-CML) & pkg\_resources & Package Management & 3 & 102 & 4.8 & 1.17$\times$ & 1.05$\times$ & 1.24$\times$ & 1.07$\times$ \\
Model-training (FWB-MT) & SciPy & Machine Learning & 5 & 1307 & 8.16 & 1.21$\times$ & 1.09$\times$ & 1.20$\times$ & 1.09$\times$ \\
Model-serving (FWB-MS) & SciPy & Machine Learning & 16 & 1463 & 7.97 & 1.23$\times$ & 1.10$\times$ & 1.22$\times$ & 1.10$\times$ \\\hline
\multicolumn{10}{c}{\textbf{Real-World Applications}} \\ \hline
OCRmyPDF & pdfminer & Document Processing & 20 & 586 & 6.4 & 1.42$\times$ & 1.19$\times$ & 1.63$\times$ & 1.00$\times$ \\
CVE-bin-tool & xmlschema & Security & 6 & 760 & 6.15 & 1.27$\times$ & 1.20$\times$ & 1.08$\times$ & 1.01$\times$ \\
Sensor-telemetry-data (SensorTD) & Prophet & IoT Predictive Analysis & 5 & 777 & 5.9 & 1.99$\times$ & 1.09$\times$ & 1.83$\times$ & 1.10$\times$ \\
Heart-Failure-prediction (HFP) & SciPy & Health Care & 5 & 982 & 8.79 & 1.38$\times$ & 1.30$\times$ & 1.46$\times$ & 1.39$\times$ \\\hline
\end{tabular}
\caption{Summary of performance improvement}
\vspace{-3mm}
\label{table:summary}
\end{table*}

\paragraph{\textbf{Experimental setup}}
We evaluate \tool{} on the AWS Lambda service. All the serverless applications are deployed as AWS Lambda functions, using the Python 3.9 runtime, with a default memory size of 1024MB, a storage size of 1024MB, and a timeout of 60 seconds. Given the diverse use cases of the multiple applications evaluated, some applications also utilize other AWS-managed services such as S3 and Amazon Elastic Container Registry (ECR). Following common practice, each application's dependencies and source code are packaged into a zip file, which is then used to create functions by transmitting the zip from an S3 bucket. Applications that use large Python libraries, such as \texttt{Pytorch}, \texttt{torchvision}, and \texttt{cve-bin-tool}, are deployed using Docker container images through AWS ECR with different memory and timeout configurations. Function URLs are created for each AWS Lambda function, and the appropriate roles are configured.

To assess the effectiveness of \tool{}, we examine 22 serverless applications. This includes four real-world applications: CVE-bin-tool~\cite{intel2024cvebin}, OCRmyPDF~\cite{ocrmypdf2024github}, Sensor telemetry data analysis~\cite{constable2024environmental}, and Heart failure prediction~\cite{sakhiya2024heart}. The remaining applications are selected from three popular benchmarks: RainbowCake~\cite{yu2024rainbowcake}, Faaslight~\cite{liu2023faaslight}, and FaaSWorkbench~\cite{kim2019functionbench}. For FaaSLight, we focus on meaningful evaluations by reporting only five representative applications that exhibited over 10\% initialization overhead, excluding trivial cases and those with minimal impact on end-to-end latency.

\paragraph{\textbf{Evaluation methodology}}
To evaluate \tool{}, we measure its effectiveness and performance across serverless applications by recording key metrics such as initialization latency, end-to-end latency, and peak memory usage. Each application is executed with 500 cold starts, and the results are averaged over five iterative runs for accuracy. The evaluation begins with the unmodified applications, which serve as the baseline. Applications with less than 10\% initialization overhead are excluded from further analysis to focus on those with significant cold-start delays. For the remaining applications, \tool{} runs its dynamic profiler to identify inefficiencies and generates optimization reports. Based on these reports, the automated code optimizer transforms the applications by replacing inefficient global imports with deferred imports. Finally, the optimized applications are redeployed, and the same metrics are collected to quantify performance improvements.

To report the results, we consider both the average and the 99\textsuperscript{th}-percentile latency as performance metrics. Reporting the 99\textsuperscript{th}-percentile latency is important because it captures the performance experienced by the slowest 1\% of requests, providing insights into potential outliers and bottlenecks. By analyzing this metric, we gain deeper insights into an application's performance under varying loads, which is critical for meeting Service Level Agreements (SLAs) and ensuring reliability. Furthermore, since serverless applications share resources in a cloud environment, optimizing memory usage improves overall resource utilization and reduces infrastructure overhead. Therefore, we also evaluate and report the memory optimizations achieved through our approach.

\begin{figure}
\setlength{\belowcaptionskip}{-15pt}
    \includegraphics[width=\linewidth]{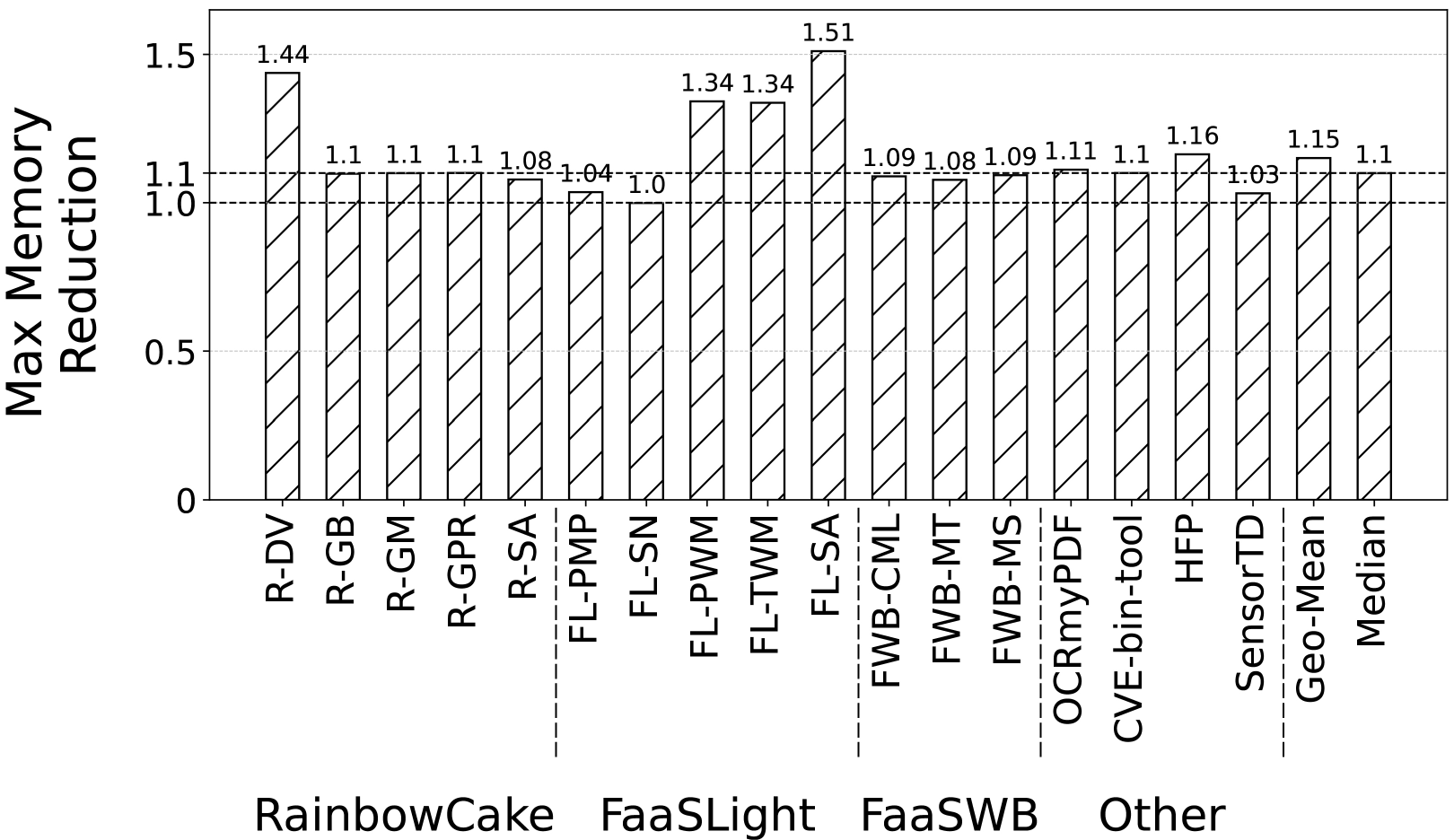}
    \caption{Memory reduction achieved by \tool{}.}
    \label{fig:mem_optimization}
\end{figure}

\paragraph{\textbf{Summary of evaluation}}
To address evaluation question \textbf{Q1}, Table~\ref{table:summary} summarizes the inefficiencies detected by \tool{} along with the optimizations achieved after addressing them. Guided by \tool{}-Profiler, our evaluation identifies inefficiencies in 17 out of 22 serverless applications evaluated. The \tool{}-guided code optimizations lead to improvements of up to 2.30$\times$ in initialization latency and 2.26$\times$ in end-to-end latency. At the 99\textsuperscript{th}-percentile, where performance bottlenecks are most pronounced, we observe improvements of up to 2.15$\times$, demonstrating \tool{}'s effectiveness in reducing tail latencies, which are critical for meeting SLA requirements. Applications with large dependency graphs, such as FL-TWM (1385 modules, avg. depth 7.57) and FWB-MS (1463 modules, avg. depth 7.97), exhibited significant speedups. This demonstrates \tool{}'s ability to scale effectively with increasing application complexity and deep call dependencies. Additionally, figure~\ref{fig:mem_optimization} details the memory reductions achieved through these optimizations, showing up to a 1.51$\times$ reduction in memory usage.

\paragraph{\textbf{\tool{} vs FaaSLight}}

To assess the effectiveness of \tool{}'s dynamic analysis and optimization, we compare its performance improvements with the speedups reported by FaaSLight, a static analysis-based approach, addressing \textbf{Q2}. Since we are unable to execute the optimized FaaSLight applications directly, the comparison relies on the performance data presented in the FaaSLight paper~\cite{liu2023faaslight}. Table~\ref{table:comparison} presents a side-by-side comparison of both tools. The results demonstrate that while FaaSLight achieves notable improvements using static analysis, \tool{}'s dynamic profiling enables greater speedups in total response latency and further reductions in memory usage for most applications. For example, in the sentiment analysis application (App11), \tool{} achieves a 2.01$\times$ improvement in total response latency compared to FaaSLight's 1.41$\times$, along with a 1.51$\times$ reduction in memory usage. These results highlight the advantages of \tool{}'s dynamic profiling in identifying and eliminating inefficiencies that static analysis approaches like FaaSLight may overlook, confirming the observation discussed in Section~\ref{staticvsDynamic}.

\begin{table}[h!]
\centering
\fontsize{7pt}{9pt}\selectfont
\begin{tabular}{@{}ccccc@{}}
\hline
\textbf{App ID} & \textbf{Tool} & \textbf{Version} & \begin{tabular}[c]{@{}c@{}}\textbf{Runtime} \\ \textbf{memory (MB)}\end{tabular} & \begin{tabular}[c]{@{}c@{}}\textbf{End-to-End} \\ \textbf{latency (ms)}\end{tabular} \\ \hline

\multirow{4}{*}{\begin{tabular}[c]{@{}c@{}}App4 \\scikit \\assign\end{tabular}} & \multirow{2}{*}{\begin{tabular}[c]{@{}c@{}}FaasLight \\ (Reported)\end{tabular}} & before & 142 & 4,534.38 \\ 
 &  & after & 140 ($1.01\times$) & 4,004.10 ($1.13\times$) \\ \cline{2-5}
 & \multirow{2}{*}{\begin{tabular}[c]{@{}c@{}}\tool{} \\ (Measured)\end{tabular}} & before & 123.64 & 3,184.67 \\ 
 &  & after & 119.38 ($1.04\times$) & 2,452.59 ($1.30\times$) \\ \hline

\multirow{4}{*}{\begin{tabular}[c]{@{}c@{}}App7 \\skimage \\lambda\end{tabular}} & \multirow{2}{*}{\begin{tabular}[c]{@{}c@{}}FaasLight \\ (Reported)\end{tabular}} & before & 228 & 7,165.54 \\ 
 &  & after & 130 ($1.75\times$) & 4,152.73 ($1.73\times$) \\ \cline{2-5}
 & \multirow{2}{*}{\begin{tabular}[c]{@{}c@{}}\tool{} \\ (Measured)\end{tabular}} & before & 112.09 & 1,821.73 \\ 
 &  & after & 112.21 ($1.00\times$) & 1,342.48 ($1.36\times$) \\ \hline

\multirow{4}{*}{\begin{tabular}[c]{@{}c@{}}App9 \\train wine \\ml-lambda\end{tabular}} & \multirow{2}{*}{\begin{tabular}[c]{@{}c@{}}FaasLight \\ (Reported)\end{tabular}} & before & 230 & 9,035.39 \\ 
 &  & after & 216 ($1.06\times$) & 7,470.49 ($1.21\times$) \\ \cline{2-5}
 & \multirow{2}{*}{\begin{tabular}[c]{@{}c@{}}\tool{} \\ (Measured)\end{tabular}} & before & 251.91 & 5,154.34 \\ 
 &  & after & 187.76 ($1.34\times$) & 3,059.18 ($1.68\times$) \\ \hline

\multirow{4}{*}{\begin{tabular}[c]{@{}c@{}}App9 \\predict wine \\ml-lambda\end{tabular}} & \multirow{2}{*}{\begin{tabular}[c]{@{}c@{}}FaasLight \\ (Reported)\end{tabular}} & before & 230 & 8,291.80 \\ 
 &  & after & 215 ($1.07\times$) & 7,071.03 ($1.17\times$) \\ \cline{2-5}
 & \multirow{2}{*}{\begin{tabular}[c]{@{}c@{}}\tool{} \\ (Measured)\end{tabular}} & before & 252.08 & 6,201.17 \\ 
 &  & after & 188.57 ($1.34\times$) & 4,147.68 ($1.50\times$) \\ \hline

\multirow{4}{*}{\begin{tabular}[c]{@{}c@{}}App11 \\sentiment \\analysis\end{tabular}} & \multirow{2}{*}{\begin{tabular}[c]{@{}c@{}}FaasLight \\ (Reported)\end{tabular}} & before & 182 & 5,551.03 \\ 
 &  & after & 141 ($1.29\times$) & 3,934.31 ($1.41\times$) \\ \cline{2-5}
 & \multirow{2}{*}{\begin{tabular}[c]{@{}c@{}}\tool{} \\ (Measured)\end{tabular}} & before & 203.54 & 4,331.43 \\ 
 &  & after & 134.72 ($1.51\times$) & 2,155.61 ($2.01\times$) \\ \hline
\end{tabular}
\caption{Comparison of \tool{} (Measured) vs FaasLight (Reported) metrics}
\label{table:comparison}
\end{table}

\paragraph{\textbf{\tool{}-Profiler overhead}}
\label{overhead}

\begin{figure}
    \includegraphics[width=1\linewidth]{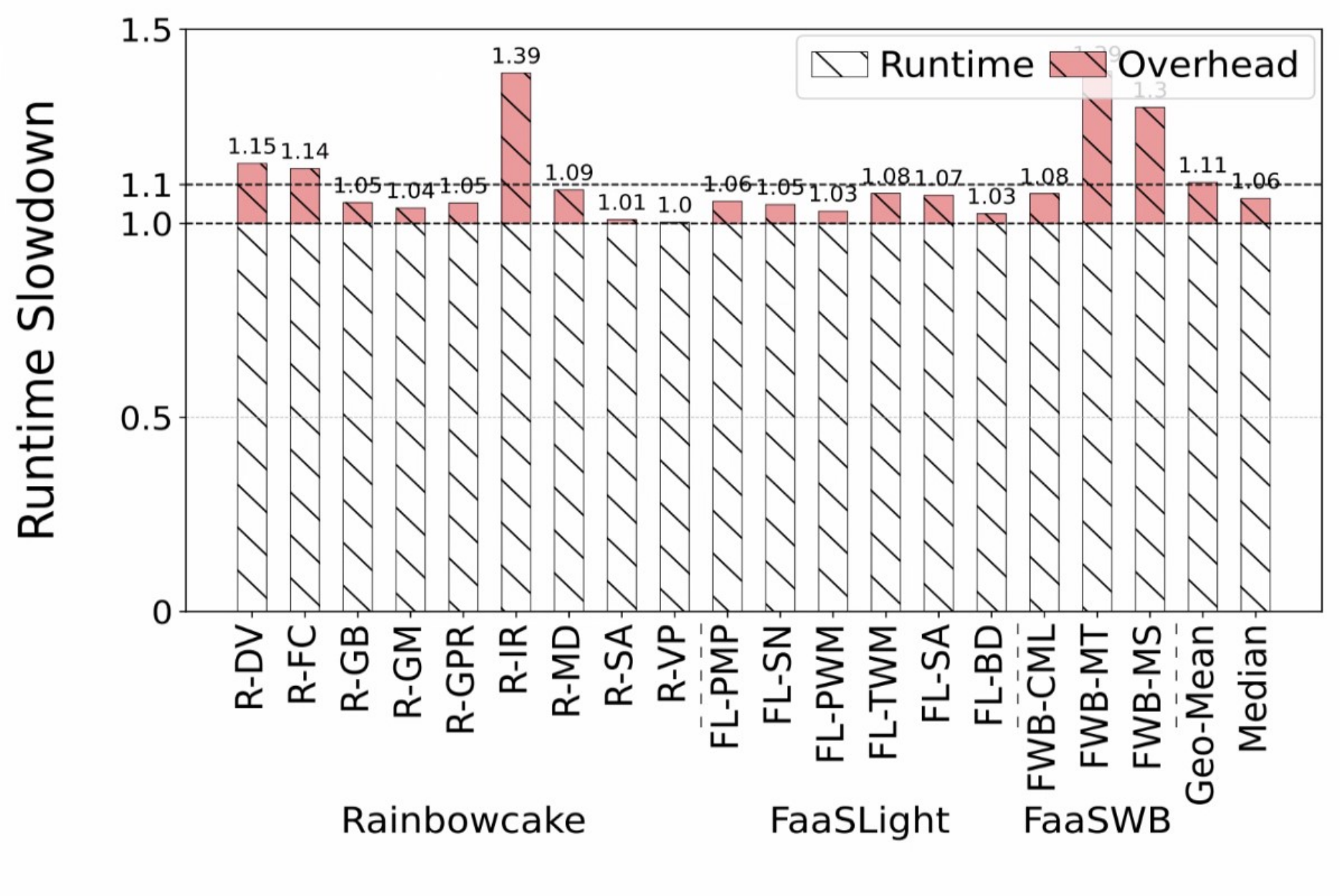}
    \caption{Runtime overhead of \tool{}.}
    \label{fig:overhead}
\end{figure}

To quantify the overhead incurred by \tool{} and answer \textbf{Q3}, we assess the runtime overhead of \tool{} on 18 Python serverless applications across all three benchmark suites, including RainbowCake, FaaSLight, and FaaS Workbench applications. During the assessment, each experiment is executed five times with 500 concurrent requests, comparing the runtime ratio with and without \tool{}-Profiler enabled. The evaluation results are illustrated in Figure~\ref{fig:overhead}. The figure shows that most serverless applications experience a maximum overhead of 10\%. Additionally, \tool{} provides an API to configure the sampling rate.

\paragraph{\textbf{Evaluation of Adaptive Profiling}}

\begin{figure}
    \setlength{\belowcaptionskip}{-10pt}
    \includegraphics[width=\linewidth]{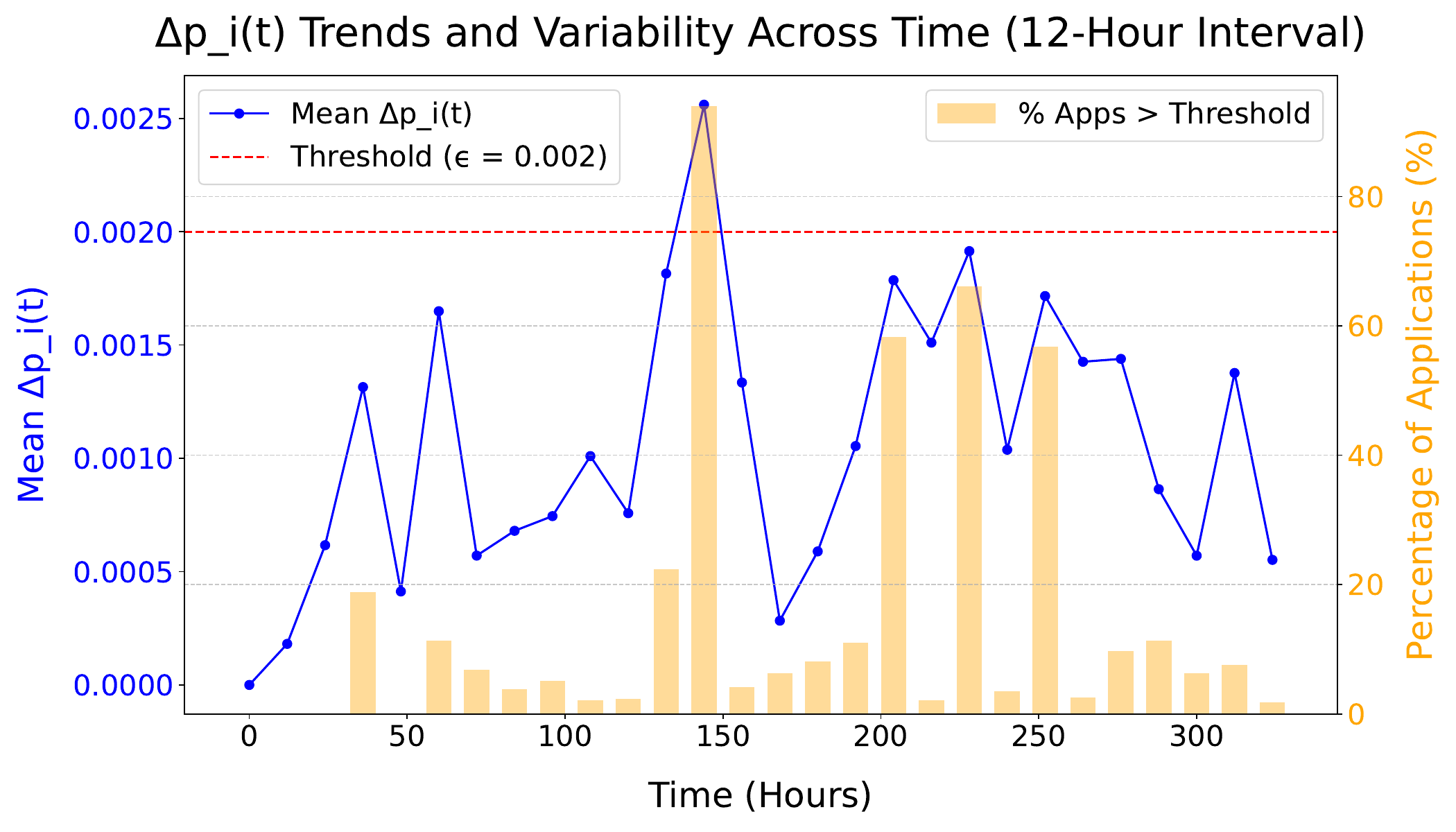}
    \caption{Trends in mean $\Delta p_i(t)$ and percentage of applications exceeding the threshold $\epsilon = 0.002$ at a 12-hour interval. The dashed line represents the threshold $\epsilon$, distinguishing stable workloads from significant workload shifts. Peaks in the percentage of applications exceeding $\epsilon$ highlight intervals where adaptive profiling is necessary.}
    \label{fig:delta_p_trends}
\end{figure}

To evaluate the effectiveness of the adaptive profiling mechanism in balancing overhead and sensitivity, we analyze empirical data from production traces. 
Figure~\ref{fig:delta_p_trends} shows trends in mean $\Delta p_i(t)$ and the percentage of applications exceeding the threshold $\epsilon = 0.002$ at a 12-hour interval. Most applications exhibit stable workloads, as reflected in low mean $\Delta p_i(t)$ values across time. Profiling during these intervals would introduce unnecessary overhead.

Peaks in the percentage of applications exceeding $\epsilon$ occur at specific time intervals (e.g., 144 hours, 228 hours), indicating significant workload shifts. These moments warrant adaptive profiling to capture meaningful changes. The selected threshold ($\epsilon = 0.002$) effectively distinguishes stable workloads from dynamic ones, enabling efficient profiling with minimal overhead. The 12-hour interval balances granularity and simplicity, ensuring that profiling focuses on critical changes while avoiding excessive sensitivity to short-term noise.

%% file: sections/7-Case-studies.tex
\section{Case Studies}
This section presents case studies that detail \tool{}'s ability to provide actionable insights at the code level.

\subsubsection{\textbf{Sentiment Analysis (R-SA)}}

\texttt{R-SA} is a sentiment analysis application featured in the RainbowCake benchmark. This scientific tool processes human-readable text, analyzes its sentiment, and classifies the text as positive, negative, or neutral. After integrating \tool{}, we upload the \texttt{R-SA} application package and its dependencies to an AWS S3 bucket. Then, we deploy \texttt{R-SA} as a Lambda function by transmitting the source code from S3. The Lambda handler reads the text input and utilizes version 3.8.1 of the natural language toolkit \texttt{nltk} library for tokenization and the \texttt{TextBlob} library for sentiment analysis. \texttt{R-SA} reports the sentiment analysis results in terms of subjectivity and polarity scores. To generate the inefficiency report, we execute 500 concurrent requests to the corresponding \texttt{R-SA} Lambda function.

\noindent{\textbf{\tool{}-Profiler Insight:}}
Table~\ref{table:rsa_report} presents the inefficiency report for the \texttt{R-SA} application generated by \tool{}. The analysis revealed that the \texttt{nltk} library contributes to approximately 70\% of the initialization latency despite having a low utilization of only 5.33\%. Further investigation identified sub-modules \texttt{sem, stem, parse, tag} as key contributors, adding 26\% to the initialization time while being unused by \texttt{R-SA}.

\noindent{\textbf{The Optimization:}}
Guided by \tool{}'s report, the automatic code optimizer transformed the \texttt{R-SA} application by applying lazy loading to the unused sub-modules \texttt{nltk.sem, nltk.stem, nltk.parse, nltk.tag}. This optimization reduced initialization latency by 1.35$\times$ and improved end-to-end latency by 1.33$\times$. Additionally, it reduced the application's peak memory usage by 1.07$\times$.

\begin{table}[h]
\centering
\begin{tabular}{cllrr}
\hline
\rowcolor{gray!30}
\multicolumn{5}{|c|}{\textbf{\tool{} Summary}} \\ \hline
\multicolumn{5}{|l|}{\textbf{Application:} rainbowcake\_sentiment\_analysis.json} \\ \hline
\rowcolor{gray!30}
\multicolumn{1}{|l|}{} & \multicolumn{1}{l|}{\textbf{Package}} & \multicolumn{1}{l|}{\textbf{Util.}} & \multicolumn{1}{c|}{\textbf{\begin{tabular}[c]{@{}c@{}}Init.\\ Overhead\end{tabular}}} & \multicolumn{1}{r|}{\textbf{File}} \\ \hline
\rowcolor{red!10}
\multicolumn{1}{|c|}{-} & \multicolumn{1}{l|}{nltk} & \multicolumn{1}{r|}{5.33} & \multicolumn{1}{r|}{69.93} & \multicolumn{1}{r|}{../nltk/\_\_init\_\_.py} \\ \hline
\rowcolor{red!30}
\multicolumn{1}{|c|}{+} & \multicolumn{1}{l|}{nltk.sem} & \multicolumn{1}{r|}{0} & \multicolumn{1}{r|}{8.25} & \multicolumn{1}{r|}{nltk/sem/\_\_init\_\_.py} \\ \hline
\multicolumn{1}{|c|}{} & \multicolumn{1}{l|}{...} & \multicolumn{1}{r|}{...} & \multicolumn{1}{r|}{...} & \multicolumn{1}{r|}{...} \\ \hline
\rowcolor{gray!30}
\multicolumn{5}{|c|}{\textbf{Call Path}} \\ \hline
\rowcolor{gray!30}
\multicolumn{1}{|l|}{\textbf{}} & \multicolumn{1}{l|}{\textbf{Package}} & \multicolumn{3}{l|}{\textbf{Path}} \\ \hline
\multicolumn{1}{|l|}{-} & \multicolumn{1}{l|}{nltk.sem} & \multicolumn{3}{l|}{\begin{tabular}[c]{@{}l@{}}handler.py:2\\ \quad$\rightarrow$ nltk/\_\_init\_\_.py:147\\
\quad\quad$\rightarrow$ $<$... parent path ...$>$\\
\quad\quad\quad$\rightarrow$ nltk/sem/\_\_init\_\_.py:44\end{tabular}} \\ \hline
\multicolumn{1}{l}{} &  &  & \multicolumn{1}{l}{} & \multicolumn{1}{l}{}
\end{tabular}
\caption{\tool{} report on Sentiment Analysis (R-SA)}
\label{table:rsa_report}
\end{table}
\vspace{-0.2 cm}

\subsubsection{\textbf{CVE Binary Analyzer}} The \texttt{CVE Binary Analyzer}, developed by Intel, is a widely used vulnerability detection tool that supports 360 checkers and 11 language-specific modules, making it well-suited for automated CVE scanning in CI/CD pipelines. For evaluation, we deployed the tool as an AWS Lambda function and integrated it with \tool{} to analyze 342 popular GitHub repositories (up to 100 KB each) using the \texttt{OpenSSL} checker. The setup scanned for \texttt{OpenSSL} versions and cross-referenced them with known CVEs. To test scalability, 500 concurrent requests were issued, allowing \tool{} to profile Python library usage during execution.

\noindent{\textbf{\tool{} Insight:}}
Table~\ref{table:cvebin_report} summarizes \tool{}'s findings, listing imported libraries along with their initialization overhead and utilization. \tool{} identifies the \texttt{xmlschema} library as responsible for 8.27\% of initialization latency, despite its low utilization of only 0.78\%. A deeper analysis shows that \texttt{xmlschema} is only required when the application processes XML files containing a Software Bill of Materials (SBOM).

\noindent{\textbf{The Optimization:}}
Using lazy loading for the \texttt{xmlschema} library, \tool{} eliminates its initialization unless explicitly required. This optimization reduces cascading initialization overhead, achieving a 1.27$\times$ improvement in initialization latency and a 1.20$\times$ improvement in end-to-end latency. The optimization also reduces maximum memory usage by 1.21$\times$ by avoiding unnecessary module loads.

\begin{table}[h]
\setlength{\belowcaptionskip}{-2pt}
\centering
\begin{tabular}{cllrr}
\hline
\rowcolor{gray!30}
\multicolumn{5}{|c|}{\textbf{\tool{} Summary}} \\ \hline
\multicolumn{5}{|l|}{\textbf{Application:} cve\_binary\_analyzer} \\ \hline
\rowcolor{gray!30}
\multicolumn{1}{|l|}{} & \multicolumn{1}{l|}{\textbf{Package}} & \multicolumn{1}{l|}{\textbf{Util.}} & \multicolumn{1}{c|}{\textbf{\begin{tabular}[c]{@{}c@{}}Init.\\ Overhead\end{tabular}}} & \multicolumn{1}{r|}{\textbf{File}} \\ \hline
\rowcolor{red!30}
\multicolumn{1}{|c|}{+} & \multicolumn{1}{l|}{xmlschema} & \multicolumn{1}{r|}{0.78} & \multicolumn{1}{r|}{8.27} & \multicolumn{1}{r|}{../xmlschema/\_\_init\_\_.py} \\ \hline
\multicolumn{1}{|c|}{+} & \multicolumn{1}{l|}{elementpath} & \multicolumn{1}{r|}{1.48} & \multicolumn{1}{r|}{8.17} & \multicolumn{1}{r|}{../elementpath/\_\_init\_\_.py} \\ \hline
\multicolumn{1}{|c|}{} & \multicolumn{1}{l|}{...} & \multicolumn{1}{r|}{...} & \multicolumn{1}{r|}{...} & \multicolumn{1}{r|}{...} \\ \hline
\rowcolor{gray!30}
\multicolumn{5}{|c|}{\textbf{Call Path}} \\ \hline
\rowcolor{gray!30}
\multicolumn{1}{|l|}{\textbf{}} & \multicolumn{1}{l|}{\textbf{Package}} & \multicolumn{3}{l|}{\textbf{Path}} \\ \hline
\multicolumn{1}{|l|}{-} & \multicolumn{1}{l|}{xmlschema} & \multicolumn{3}{l|}{\begin{tabular}[c]{@{}l@{}}handler.py:11\\ \quad$\rightarrow$ cve\_bin\_tool/cli.py:71\\ \quad\quad$\rightarrow$ cve\_bin\_tool/sbom\_detection.py:8\\ 
\quad\quad\quad$\rightarrow$ cve\_bin\_tool/validator.py:11\end{tabular}} \\ \hline
\multicolumn{1}{l}{} &  &  & \multicolumn{1}{l}{} & \multicolumn{1}{l}{}
\end{tabular}
\vspace{-3mm}
\caption{\tool{} report on CVE binary analyzer}
\vspace{-3mm}
\label{table:cvebin_report}
\end{table}

%% file: sections/3-Related-work.tex
\section{Related work}
\label{relatedwork}
\noindent{\textbf{Platform-Level Runtime Optimizations:}} Several techniques have been proposed to enhance infrastructure efficiency and mitigate cold start latency through optimized resource allocation and scheduling of serverless functions. These methodologies encompass shared resource utilization~\cite{li2022help}, automatic memory deduplication~\cite{saxena2022memory}, function caching~\cite{chen2023s}, compression~\cite{basu2024codecrunch}, advanced scheduling algorithms~\cite{pan2023sustainable}, and the reuse of pre-warmed instances~\cite{bhasi2021kraken, gunasekaran2020fifer, roy2022icebreaker, shahrad2020serverless}. Additional approaches focus on proactively loading libraries into warm containers to reduce the cold start overhead~\cite{sui2024pre}. \textit{While effective at the platform level, these approaches leave application-level inefficiencies, such as suboptimal library usage, unaddressed.}

\noindent{\textbf{User-Directed Serverless Runtime Optimizations:}} User-directed optimizations involve configuring serverless runtime policies to reduce cold start times. Techniques include checkpointing~\cite{ao2022faasnap, du2020catalyzer, silva2020prebaking} to save function state, provisioned concurrency~\cite{provisionedConcurrencyAWS} to keep instances warm, adjusting memory~\cite{improveColdstartByIncreasingMemory} and compute resources~\cite{optimisingServerlessForBBC} to optimize performance, keep-alive~\cite{fuerst2021faascache, pan2022retention, roy2022icebreaker, shahrad2020serverless} configurations to prevent premature termination, and layering dependencies~\cite{yu2024rainbowcake} to reduce loading overhead by caching and updating them independently. \textit{However, these runtime-level policies lack the granularity required to address code-level inefficiencies, such as unused or infrequently used libraries.}

\noindent{\textbf{Code-level optimizations:}} 
Code-level techniques aim to reduce initialization time and improve application performance by code optimization. Examples include function fusion to minimize initialization overhead~\cite{lee2021mitigating}, function decomposition into smaller units~\cite{kalia2021mono2micro, nitin2022cargo, abgaz2023decomposition}, and serverless function compression~\cite{liu2023faaslight}. General-purpose tools like JAX~\cite{frostig2018compiling}, GraalVM~\cite{graalvm}, ProGuard~\cite{proguard}, and R8~\cite{r8_android} use static analysis to optimize runtime performance. \textit{However, these tools do not adapt to dynamic runtime behavior, limiting their effectiveness in serverless workloads with varying library usage patterns.}

Unlike prior approaches that overlook application context or dynamic behavior, \tool{} leverages runtime profiling of the serverless application to observe real-time library usage patterns, capturing dynamic dependencies and workload-specific inefficiencies.

%% file: sections/9-Conclusions.tex
\section{Conclusions}
\label{conclusions}
This paper introduces \tool{}, a profile-guided optimization tool that identifies and mitigates workload-dependent library inefficiencies. By combining runtime profiling, automated optimization, and adaptive monitoring, \tool{} addresses gaps in static analysis. Evaluation results show up to 2.30$\times$ faster initialization, 2.26$\times$ improved end-to-end latency, and 1.51$\times$ lower memory usage. These outcomes highlight the value of runtime profiling in optimizing serverless applications for enhanced resource use and responsiveness. In the future, we will extend \tool{} to support other popular serverless programming languages, such as JavaScript.


%% file: sections/11-references.tex
\bibliographystyle{IEEEtran}
\bibliography{BibFile}